\documentclass{aa}  

\usepackage{graphicx}
\usepackage{txfonts}
\usepackage{hyperref}
\hypersetup{colorlinks=true,
            linkcolor=blue,
            urlcolor=blue,
            citecolor=blue}
\usepackage{amsmath}
\usepackage{xcolor}

\newcommand{\BW}{\textit{Blastwave}~}
\newcommand{\SB}{\textit{Superbubble}~}
\newcommand{\RT}{\textit{RT}~}

\begin{document}

   \title{The Seven Dwarfs illuminated}
   \subtitle{The impact of radiation on dwarf galaxies and their circumgalactic medium}

   \author{Bernhard Baumschlager\inst{1}
          \and
          Sijing Shen\inst{1}
         \and
          James W. Wadsley\inst{2}
          \and
          Benjamin Keller\inst{3}
          \and 
          Robert Wissing\inst{1}
          \and 
          Lucio Mayer \inst{4} 
          \and 
          Piero Madau \inst{5}
          \and 
          Ferah Munshi \inst{6}
          \and 
          Alyson Brooks \inst{7}
          }
   \authorrunning{B. Baumschlager et al.}

   \institute{
             Institute of Theoretical Astrophysics, University of Oslo, P.O. Box 1029 Blindern, N-0315 Oslo, Norway
              \email{bernhard.baumschlager@astro.uio.no}
        \and
             Department of Physics and Astronomy, McMaster University,
             Hamilton, L8S 4M1, Canada 
        \and 
            Department of Physics and Material Science, University of Menphis, 3720 Alumni Avenue, Memphis, TN 38152, USA
       \and 
           Department of Astrophysics, University of Zurich, Winterthurerstrasse 190, CH-8057 Zurich, Switzerland
         \and  
            Department of Astronomy \& Astrophysics, University of California, 1156 High Street, Santa Cruz, CA 95064, USA
        \and 
            Department of Physics \& Astronomy, George Mason University, 4400 University Drive, Fairfax, VA 22030, USA   
        \and
            Department of Physics and Astronomy, Rutgers University, the State University of New Jersey, 136 Frelinghuysen Road, Piscataway, NJ 08854-8019, USA
             }

   \date{}

 
  \abstract
  {We present a high-resolution cosmological zoom-in simulation of a group of field dwarf galaxies which includes on-the-fly radiative transfer (RT) and is evolved to $z=0$. Emission from stars is included according to age-dependent spectral energy distributions, and a redshift-dependent UV background (UVB) is modelled by background particles and propagated through the simulation volume. The simulation also incorporates star formation, supernova (SN) feedback, a non-equilibrium primordial chemical network, and metal cooling and diffusion.
  We compare results from this simulation to previous simulations with the same initial conditions and largely similar galaxy formation physics, but without RT. 
  The inclusion of RT results in the formation of eight additional faint dwarf galaxies with stellar masses of $10^{4}\, \rm M_{\odot}$ to several $10^{5}\, \rm M_{\odot}$ and only old stellar populations, similar to the observed Ultra-Faint Dwarf galaxies (UFDs). They formed before and during cosmic reionisation and were mostly quenched by $z \sim 3-4$ when the reionisation was complete. The simulated galaxies follow many observed scaling relations such as the stellar mass-halo mass relation, the mass-size relation, and the luminosity-velocity dispersion relation. However, it appears to underpredict the stellar iron abundances for the faintest population with $L_{v}\, < 10^{6} \rm L_{\odot}$. 
  For the more massive "classic" dwarf galaxies, radiative feedback suppresses star formation, making it less bursty and reducing explosive outflows.
  This consequently reduces the dark matter core sizes by a factor of 2-3, rendering the core sizes ($\sim$ 1 kpc) more consistent with observations.
  The distribution of neutral hydrogen \ion{H}{I} in the circumgalactic medium (CGM) is  ubiquitous with a covering fraction of unity within $R_{vir}$, in good agreement with observations. It is rather insensitive to radiative or SN feedback at $z=0$, but at $z>5$ it is much higher in the RT simulation. 
  In contrast, the distribution of low ions like \ion{Si}{II} is very compact and declines sharply beyond the ISM scale. \ion{C}{IV} and \ion{O}{VI} have a more extended distribution, but their column densities are generally below the detection limit. Radiative feedback leads to smaller column densities of the metal ions, partly due to the reduction of total metal production, and partly because hard photons from the stellar radiation escape the ISM and further ionise the CGM. The abundance of \ion{C}{IV} is particularly sensitive to the latter effect. 
  }


   \maketitle
%
\section{Introduction}\label{sec:intro}

Dwarf galaxies constitute the low mass end of the galaxy zoo, and are the most numerous and least luminous systems in the Universe.
In the framework of hierarchical structure formation of the $\Lambda$CDM paradigm, dwarf galaxies form within the earliest collapsing halos, and become the building blocks for more massive galaxies.
Observations of dwarf galaxies have posed significant challenges to the $\Lambda$CDM model. 
First, simulations predict that Milky Way-mass halos should host significantly more satellite galaxies than observed around the Milky Way and M31. This discrepancy, known as the ``missing-satellite problem,'' suggests that either the predicted small halos fail to form stars or that many remain undetected \citep[e.g.][]{moore99,klypin99}.
Second, dark matter (DM)-only simulations predict a cuspy inner dark matter profile, whereas observations of rotation curves of dwarf galaxies suggest that the dark matter profile is cored, i.e. has a near-constant inner slope - the so-called "cusp-core problem" \citep[e.g.][]{flores94,navarro96,deblok02}.
Finally, the most massive subhalos in dark matter-only simulations of Milky Way-sized galaxies are too massive compared to observed Milky Way satellites; this is referred to as the "too-big-to-fail (TBTF) problem" \citep{boylankolchin11,boylankolchin12}. 

Although alternative dark matter models are invoked to alleviate these tensions \citep[e.g.][]{Vogelsberger14, Fitts19, Mina20, Nori24, Shen24}, many studies have shown that baryonic processes can also have a significant impact. Due to their shallow gravitational potential wells, dwarf galaxies are sensitive to mechanical and radiative feedback from stars and potentially active galactic nuclei (AGN). After reionisation, photoionisation heating from the cosmic ultraviolet background (UVB) suppresses gas accretion into the small halos \citep[e.g.][]{Efstathiou92,Bullock2000, Dijkstra04, madau08, Gutcke22b}, which alleviates the "missing satellite problem". Galactic outflows driven by stellar feedback can rapidly change the central gravitational potential wells, leading to cusp-core transformation \citep[e.g.][]{Mashchenko08, Governato2010,Pontzen2012,madau14,fitts17,Mostow24}. The cored DM profile reduces the peak velocity in galaxy rotation curves, which alleviates the TBTF problem \citep{Zolotov12, Brooks14, Wetzel16}. 

Meanwhile, the inclusion of feedback is also crucial for modern hydrodynamic simulations of dwarf galaxies to successfully reproduce observed properties of Local Group dwarf galaxies, such as the stellar mass content \citep[e.g.][]{behroozi13,moster13,brook14,garrison-kimmel14,read17}, star-formation history \citep[e.g.][]{brown14,weisz14}, mass-metallicity relation \citep[e.g.][]{mcconnachie12,kirby13}, and stellar/gas kinematics \citep{mcconnachie12}. One of the key processes is feedback-driven galactic outflows. Numerous studies have shown that galactic outflows are of critical importance in suppressing star formation, ejecting metals into the circumgalactic medium (CGM) and the intergalactic medium (IGM), thus lowering stellar and gas metallicity in the disc and reducing the stellar bulge-to-disc ratio \citep[e.g.][]{shen_2014, onorbe15, Wheeler15, fitts17}. Traditionally, supernova feedback is generally invoked in generating outflows, but it is still under debate whether supernova alone are responsible for reproducing all the observed properties of local dwarf galaxies. This is partly because subgrid models are often adopted for supernova feedback, which can either overestimate or underestimate the SN energy/momentum injection into the interstellar medium (ISM). Indeed, the properties of dwarf galaxies can be very sensitive to supernova feedback models, especially the outflow and CGM properties \citep[e.g.][]{mina21}, and convergence between various models is achieved only with extreme resolution \citep[e.g.][]{Smith18}. More recently, cosmological simulations, such as EDGE \citep{agertz20} and LYRA \citep{Gutcke21}, were able to push to very high resolution such that the Sedov-Taylor phase of the supernova explosion is largely resolved, and thus the momentum injection is not dependent on subgrid models. However, even in this case, results are not fully converged between different numerical methods \citep{Hu2023}. Moreover, the impact of supernova also depends on the ISM conditions and the baryonic cycle, and the interplay between supernova feedback and other physical processes including, e.g. gas accretion, turbulence, H$_{2}$ chemistry, star formation, magnetic fields, and possible AGN feedback, are nonlinear and complex \citep[e.g.][]{Hu16, Munshi19, Agertz15, Koudmani21, martin-alvarez23, Sharma23, Arjona24}. 

Radiation from stars and AGN is one of such processes that significantly impacts the condition of the ISM (and also the CGM and IGM, i.e. the entire baryonic cycle). As mentioned previously, cosmic reionisation directly affects the formation of the faintest galaxies by reducing or quenching gas accretion onto the smallest halos. At the ISM scale, detailed ISM simulations have shown that photoionising and photodissociating radiation from massive stars precedes supernova explosions and can significantly reduce the ISM densities where SNe occur, thus boosting local supernova efficiency and wind launching \citep[e.g.][]{walch15, kim17, peters17}. In addition to radiative heating, radiation pressure can directly inject momentum into the ISM and drive galactic winds itself, especially for starburst galaxies and active galaxies \citep{Fabian12}. 

For galactic scale simulations and cosmological simulations, on-the-fly radiative transfer (RT) is computationally costly because of its requirements for spatial and temporal resolutions. However, great progress has been made in recent years, and fully coupled RT has become more feasible, for example, in simulations of the Epoch of Reionisation (EoR) such as CROC \citep{Gnedin14}, SPHINX \citep{Rosdahl18} and THESAN \citep{Kannan22, Borrow23}. Although these simulations are usually stopped at a high redshift (e.g. $z \sim$ 6), the self-consistent modelling of the evolving UV background already revealed interesting results. For dwarf galaxy formation and quenching, the inhomogeneity of the ionising background and the timing of reionisation in a specific region can cause significant scatter in the star formation history in small halos \citep{Katz20}. Such an effect cannot be captured with the traditional approach in cosmological simulations, where a spatially uniform UVB is turned on instantaneously and is only a function of time. As for the interplay between radiation, SN feedback and star formation, both isolated galaxy simulations \citep{Rosdahl2015,emerick18, Deng24} and cosmological zoom-in simulations \citep{agertz20, hopkins20, martin-alvarez23, Rey2025} with full RT have shown that radiative heating suppresses global star formation and stellar mass, but it is still unclear how the global supernova efficiency is affected. While some studies \citep{Rosdahl2015, hopkins20, Deng24, Rey2025} found that radiative heating prevents the formation of dense clouds and "smooths out" the ISM, resulting in a less bursty star formation history and more gentle SN-driven outflows, others \citep[e.g.][]{emerick18} found the opposite, namely radiative feedback helps to create ISM conditions that favour supernova feedback, which is similar to what is found in small-scale ISM simulations. As rapid fluctuations of the central gravitational potential well are essential for the DM cusp-core transformation, it is expected that the burstiness of star formation will have an impact on DM core properties.  

Observationally, with the advent of digital sky surveys and deep imaging, the past two decades have seen a rapid growth of new discoveries of the faintest galaxy population, the ultrafaint dwarf galaxies (UFDs) \citep[e.g.][]{simon19,Richstein24}, typically with luminosities $L < 10^{5}$ L$_{\odot}$. These galaxies have predominantly very early star formation histories \citep[e.g.][]{brown14} and low metallicities \citep[e.g.][]{Fu23}, indicating that they are likely to be true relics of reionisation. As such, it is important to include radiative transport in simulations of UFDs. Moreover, their smaller mass and quiescent star formation activities today imply insufficient SN feedback and energy injection for the cusp-core transformation, making them ideal to break the degeneracy between alternative dark matter models and baryonic processes in the cusp-core problem \citep{governato12}. With recently launched missions such as the Euclid satellite \citep{Eucild24,Marleau24} and Rubin Observatory \citep{Ivezic19}, and future missions such as the Roman Space Telescope \citep{Akeson19} and Arrakihs \citep{Corral24}, observations will continue to push the detection limit of faint dwarf galaxies and low surface brightness (LSB) features, which will significantly improve our understanding of dark matter and galaxy formation in general. 

In addition to the stellar and ISM components, observations of the CGM can also reveal crucial information on the baryonic cycle, feedback, and environmental effects in dwarf galaxy formation. Observations of the CGM around nearby dwarf galaxies, mainly through absorption line studies (e.g. HST/COS) found that although neutral hydrogen (\ion{H}{I}) is ubiquitous, detections of the low ionisation metal species such as \ion{C}{II} and \ion{Si}{II} are generally rare around dwarf galaxies, whereas detections of intermediate or highly ionised metal lines (e.g. \ion{C}{IV}) are relatively more common but still less frequent than those around more massive galaxies \citep[e.g.][]{Prochaska11, Bordoloi14, Liang14, Burchett16, Johnson17, Zheng20, Zheng24,Mishra24}. Because observations only probe ions rather than the total gas or metal mass, it is important to understand the role of the local radiation field on the ionisation states of the CGM. For simplicity, CGM analysis of cosmological simulations usually assumes a uniform UVB radiation when post-processing simulations. While this may be a good approximation at large distances in the outer halo, it is inaccurate for the inner CGM, as local radiation can dominate the radiation field and impact ion abundances, at least for high-redshift, starburst galaxies \citep{shen13} and Milky Way type galaxies \citep{Zhu24}. It remains to be explored how the CGM of dwarf galaxies is affected by the local radiation field, and RT is an essential tool for such investigation.    

Much effort has been devoted to simulating dwarf galaxy formation and evolution. In particular, cosmological zoom-in simulations of field dwarf galaxies \citep[e.g.][]{Governato2010, shen_2014, onorbe15, Wheeler15,fitts17, jeon17, Smith18,revaz18,Fitts19,Munshi19, Wright19, agertz20,Gutcke21,mina21, martin-alvarez23, Azartash-Namin24}, and satellite dwarf galaxies around MW-like galaxies or in Local Group-like environments \citep[e.g.][]{Brooks14, Sawala16,Wetzel16, Santos-Santos18, applebaum21} have achieved great success in reproducing the observed dwarf galaxy population in the local Universe, and in advancing our understanding of the key physics that shapes galaxy formation at the faintest end. More recent studies \citep[e.g.][]{Li21, mina21, Piacitelli25} also investigated the CGM properties around dwarf galaxies at low redshift. However, full RT has only been included in such simulations more recently, for example, in EDGE \citep{agertz20, Rey2025} and Pandora \citep{martin-alvarez23}. These simulations, together with the RT EoR simulations, have shown the importance of modelling the UVB and local radiation field properly in dwarf galaxy formation. Nevertheless, due to computational cost and challenges, cosmological RT simulations often have to be terminated at a relatively high redshift, typically at $z > 3$, or focus on the smallest halos in an isolated environment, typically with halo mass $M_{h} \sim 10^{9}$ M$_{\odot}$, but observations of dwarf galaxies and the CGM are predominantly at the local Universe and for more massive systems. Therefore, it remains to be investigated how the impact of radiation evolves throughout cosmic history in various halo masses, how it interacts with star formation, supernova feedback, and how it affects cusp-core transformation and CGM ionisation states.  

In this work, we explore these questions using a cosmological zoom-in simulation of a group of field dwarf galaxies with host halo mass $M_\mathrm{vir}\lesssim10^{10.5}\,\mathrm{M}_\odot$ with on-the-fly RT evolved to $z=0$. Radiative transfer is performed with a tree-based reverse ray tracing algorithm {\sc Trevr2} \citep{wadsley23}. The initial condition is adopted from the Seven Dwarf Simulation from \citet{shen_2014}, which used the "blastwave" delayed cooling model for supernova feedback \citep{stinson06}. It successfully produced field dwarf galaxies that follow many observed scaling relations in total stellar and gas mass, metallicity, kinematics, and morphology. The simulated galaxies have extremely bursty star formation histories, which cause persistent DM cores \citep{madau14}, and impulsive galactic outflows, resulting in large metal-enriched CGM. The same group of galaxies was also simulated in \citet{mina21} with the more realistic "superbubble" feedback model \citep{keller14} without ad hoc cooling delays to explore the impact of SN feedback models on dwarf galaxy properties (see also \citet{Azartash-Namin24} for a similar study but with H$_{2}$ chemistry and H$_{2}$-based star formation). The authors found that while global quantities of galaxies with halo mass $M_{h}>10^{10}$ M$_{\odot}$ remain similar with different feedback models, the DM cores, the CGM metals, and the formation of smaller galaxies ($M_{h} \sim 10^{9}$ M$_{\odot}$ and below) are very sensitive to feedback. In this study, we adopt the exact galaxy formation physics from \citet{mina21}, but with the addition of RT, which models the transport of radiation from both the UVB and local stellar sources. The use of the same initial condition and resolution makes it ideal to compare with both previous simulations to single out the impact of radiation.

The paper is structured as follows.
In Section \ref{sec:sim} we briefly present the simulation setup and summarise the main features of the RT implementation and the treatment of the radiation field. In Section \ref{sec:results} we describe the general properties of the simulated dwarf galaxies and analyse the impact of on-the-fly RT on the formation and evolution of dwarf galaxies.
We study the properties of the DM halos of the two most massive halos in the simulation, focusing on the cusp-core transformation in Section \ref{sec:dm}.
In Section \ref{sec:cgm} we explore the CGM of the group of dwarf galaxies and compare it to observational constraints.
A summary of our findings is provided in Section \ref{sec:summary}.

\section{Simulation}\label{sec:sim}

The simulation has the same initial condition and galaxy formation physics as in \citet{mina21}, except for the on-the-fly radiative transfer (RT) models. Here, we briefly outline its general setup. The zoom-in simulation was performed using the TreeSPH code {\sc Gasoline2} \citep{wadsley17}, adopting a cosmology with $\Omega_\mathrm b=0.042$, $\Omega_\mathrm m=0.24$, $\Omega_\Lambda=0.76$, $h=0.73$ and $\sigma_8=0.77$. 
The simulation box, with a side length of 25 co-moving Mpc (cMpc), hosts a $\sim2\,\mathrm{cMpc}$ wide high resolution zoom-in region. This zoom region contains about 6 million dark matter and equally many SPH particles, with a mass resolution of $m_\mathrm{dm}=1.6\times10^4\,\mathrm{M}_\odot$ and $m_\mathrm{SPH}=3.3\times10^3\,\mathrm{M}_\odot$. The resolution of this simulation is typical for state-of-the-art zoom-in simulations of Milky Way-like galaxies and their satellites \citep[e.g.][]{wheeler19, applebaum21}, although recent simulations of field dwarf galaxies have reached higher resolution \citep[e.g.][]{agertz20, Gutcke21, munshi21}. Since the main objective of this work is to investigate the impact of radiative feedback, we choose to keep the same resolution as in previous works to minimise the uncertainties introduced by resolution.   

Star formation follows the subgrid model from \citet{stinson06}, where gas particles probabilistically convert into star particles when gas cools below $10^{4}$ K,
reaches a density threshold of $n_\mathrm{H}\ge100\,\mathrm{atoms\,cm}^{-3}$, and is in a convergent flow. 
The star formation rate follows the Schmidt law \citep{schmidt59}:
\begin{equation}\label{eq:schmidt}
    \frac{\mathrm{d}\rho_\star}{\mathrm{d}t}=\varepsilon_\mathrm{SF}\,\frac{\rho_\mathrm{gas}}{t_\mathrm{dyn}} \propto \rho_\mathrm{gas}^{1.5},
\end{equation}
where $\rho_\star$ and $\rho_\mathrm{gas}$ are the stellar and gas density, respectively, $\varepsilon_\mathrm{SF}=0.1$ is the star formation efficiency and $t_\mathrm{dyn}$ the local dynamical time.
The newly formed star particle has an initial mass equal to the parent gas particle,  $m_\star=3.3 \times10^3\,\mathrm{M_\odot}$ and each star particle represents a simple stellar population with its stellar mass distribution described by a \citet{kroupa01} initial mass function (IMF). 

During their lifetimes and at the end of their life, stars return energy, mass, and metals into their surrounding medium. We include feedback from stellar winds, Type Ia and Type II Supernova. These feedback models (both energetic and chemical) remain the same as in \citet{mina21}. In particular, we use the ``superbubble'' subgrid model from \citet{keller14} for SN II. This model takes into account the development of the superbubble from multiple SNe events, and thermal conduction between the hot and cold phases, which is an important process for cold gas evaporation \citep[e.g.][]{El-Badry19}. 

Following \citet{shen_2014}, we track oxygen and iron yields separately using the model from \citet{raiteri96}.The oxygen and iron abundances are then converted to alpha-element and iron peak-element abundances, respectively, assuming the solar abundance pattern from \citet{asplund09}. A turbulent diffusion model for metals and thermal energy is also included to account for sub-resolution turbulent mixing processes during gas advection, as described in detail in \citet{shen10}. 

The initial conditions are identical to those used in the simulations of \citet{shen_2014} and \citet{mina21}. We will compare with both previous simulations, although the main comparison will be with the \citet{mina21} data, since our goal is to understand the dependencies of dwarf galaxy properties on UVB and stellar radiation. In the following, we will refer to the simulation analysed in \citet{shen_2014} as \BW, the one analysed in \citet{mina21} as \SB, in accordance with their stellar feedback prescription, and our new run will be labeled as \RT.

\subsection{Radiative transfer}\label{sec:rt}

On-the-fly radiative transfer is solved using \textsc{Trevr2} \citep{wadsley23}, the new tree-based reverse ray tracing algorithm implemented in \textsc{Gasoline2}. 
\textsc{Trevr2} combines gas particles in tree cells according to the transmission averaging scheme and the rays are built along the direction of HEALPix \citep{gorski05} cones. 
This results in a fast RT algorithm which scales as $\mathcal{O}~(N_\mathrm{active}\,\mathrm{log}_2\left(N\right))$, where $N_\mathrm{active}$ is the number of ``active particles'' (i.e. particles whose properties will be updated in a certain time step), and $N$ is the number of sources. 

The source spectra are reconstructed using the piece-wise power-law (PWPL) method described in detail in \citet{baumschlager23}. In brief, the spectra are represented at nine photon energies located just above and below the ionisation energies of H and He, at 136 and 500 eV and one additional more classical broad FUV band (5.6-11.2 eV). The placement of the selected photon energies just above and below the ionisation potential of H and He allows us to accurately track jumps in intensity across the ionisation edges. The method assumes that the spectral shape follows a power-law between these energies. Table \ref{tab:radbands} summarises the selected photon energies and the physical effects they are used for. Furthermore, the photo-ionisation cross-sections for H and He are represented as power-laws, fitted to the photo-ionisation cross-sections from \citet{verner96} such that:
\begin{equation}\label{eq:dpl}
   \sigma_{\rm i,j}(\nu) = \nu^{\beta_{\rm i,j}} \left(S_{\rm i,j} + T_{\rm i,j}\,\nu\right),
\end{equation}
where $\beta_{\rm i,j}$ is the power-law slope for element $i$ in the radiation band $j$, i.e. for H in the HeII ionising band (54.4-500.0 eV), and $S_{\rm i,j}$ and $T_{\rm i,j}$ are fitting coefficients, provided in \citet{baumschlager23}.

\begin{table}
    \caption{Radiation bands used in the simulation}
    \centering
    \begin{tabular}{lcl}
         \hline \hline
         Band & Photon energy & Physical effect \\
              & [eV]         & \\
         \hline
         FUV  & 5.6 - 11.2 & \\
         LW-l & 11.2$^+$ & \\
         LW-h & 13.6$^-$ & \\
         HI-l & 13.6$^+$ & H ionisation/heating \\
         HI-h & 24.6$^-$ & H ionisation/heating \\
         HeI-l & 24.6$^+$ & H \& He ionisation/heating \\
         HeI-h & 54.4$^-$ & H \& He ionisation/heating \\
         HeII-l & 54.4$^+$ & H, He \& He$^+$ ionisation/heating \\
         HeII-h & 136.0 & H, He \& He$^+$ ionisation/heating \\
         XUV & 500.0 & H, He \& He$^+$ ionisation/heating \\
    \hline
    \end{tabular}
    \label{tab:radbands}
\end{table}

\subsection{Sources of radiation}

The PWPL spectral reconstruction of the radiation field allows us to combine different sources of radiation, and for the presented simulation, we include both stellar radiation and the UV background. Each stellar particle is treated as a source of radiation, with its intensity and spectral shape defined by its age. The intensity per radiation bin is pre-calculated based on \textsc{starburst99} \citep{leitherer14} spectra. 
We assume that the stellar population follows a \citet{kroupa01} IMF and use the Padova AGB stellar evolution tracks for solar metallicity. 
We consider population ages between 1 Myr and 13 Gyr for our simulation. For younger and older populations, we assume the spectral shape and intensity of the 1 Myr and 13 Gyr populations, respectively.

The UVB radiation is provided through 128 background sources evenly distributed in a spiral pattern on a sphere at the edge of the simulation box, as described in \citet{grond19,baumschlager23}.
These background sources follow the redshift dependent background spectrum of \citet{HM12} from $z\sim15.2~\mathrm{to}~0$. 
Rather than instantly switching on the UVB, we increase the intensity between redshift 16 and 15.2 from zero intensity to the \citet{HM12} intensity.

\subsection{Chemical network coupled with RT}

The simulation includes radiative cooling and heating over a temperature range $T=10-10^9\,\mathrm{K}$. The narrow band approximation and the power-law fits to the photo-ionisation cross section allow us to analytically approximate the photo-ionisation and photo-heating rates for arbitrary shapes of radiation source spectra with great accuracy \citep[cf. Equation (8) and (9) in][]{baumschlager23}. Thus, the abundances, heating and cooling rates of the primordial species (H, H+, He, He+ and He++) are computed with the non-equilibrium chemical network \citep{shen10}, with photoionisation rates calculated from both the UV background \citep{HM12} and local stellar radiation. Chemistry of molecular hydrogen ($\rm H_{2}$) is not followed in this simulation for simplicity. However, we note that $\mathrm{H}_{2}$ formation and disassociation can play an important role in dwarf galaxies \citep[e.g.][]{Munshi19}. The current implementation of H$_{2}$ in {\sc Gasoline2} and {\sc ChaNGa} is based on a local estimation of the Lyman-Werner radiation \citep{Christensen12}, and we plan to couple the solver fully with RT in an upcoming paper. For this work, it is informative to isolate and understand the impact of ionising radiation without involving the complexity of H$_{2}$-based star formation. 

For metal-line cooling and heating, we continue to use the rates from {\sc CLOUDY} \citep{ferland17} assuming photoionisation equilibrium under a uniform UVB. We note that this approach is not entirely consistent, as the ionisation abundances of metals are dependent on the local radiation field in the ISM and the CGM. However, including a metal chemical network or tabulated cooling rates with a non-uniform, multi-frequency radiation field is computationally or memory expensive for cosmological simulations evolving to $z=0$. Therefore, we have adopted this simpler model for metal cooling and defer a more detailed study to future work.     

\section{Properties of the simulated galaxies}\label{sec:results}

\begin{figure*}
    \centering
    \includegraphics[width=\columnwidth]{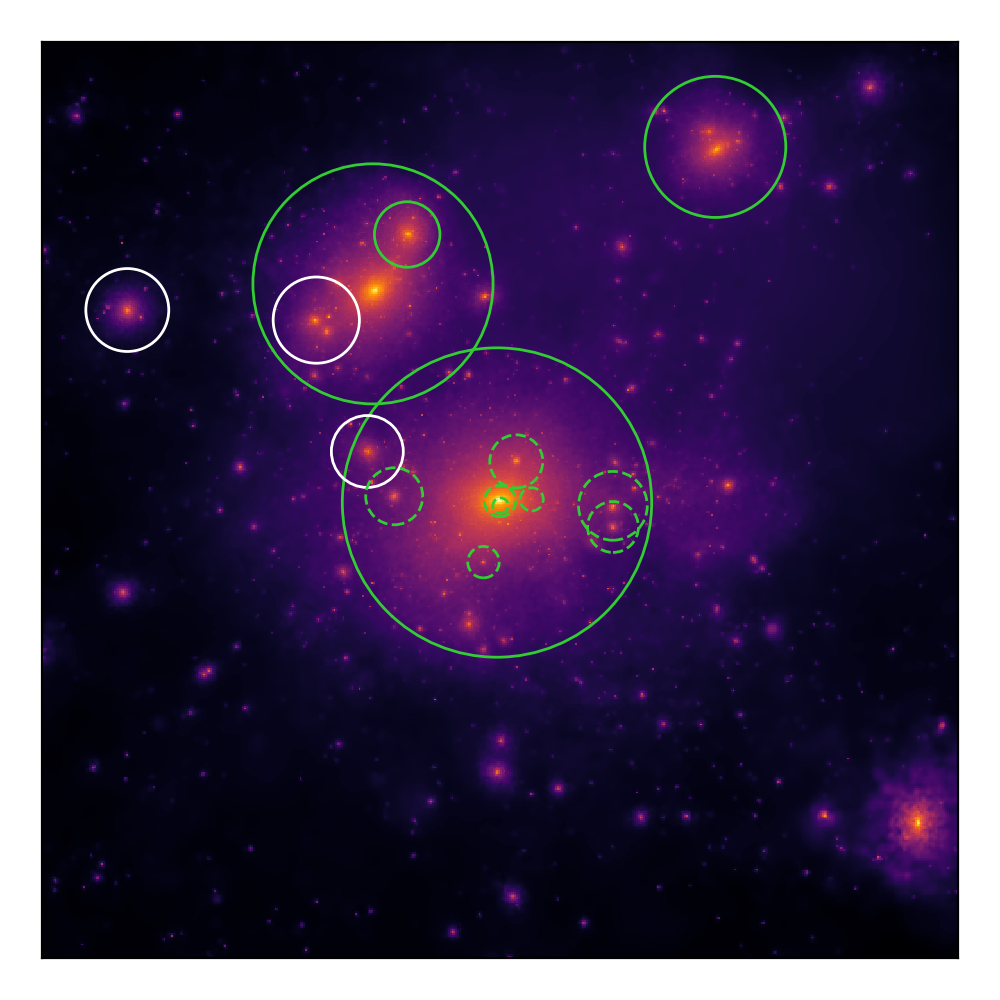}
    \includegraphics[width=\columnwidth]{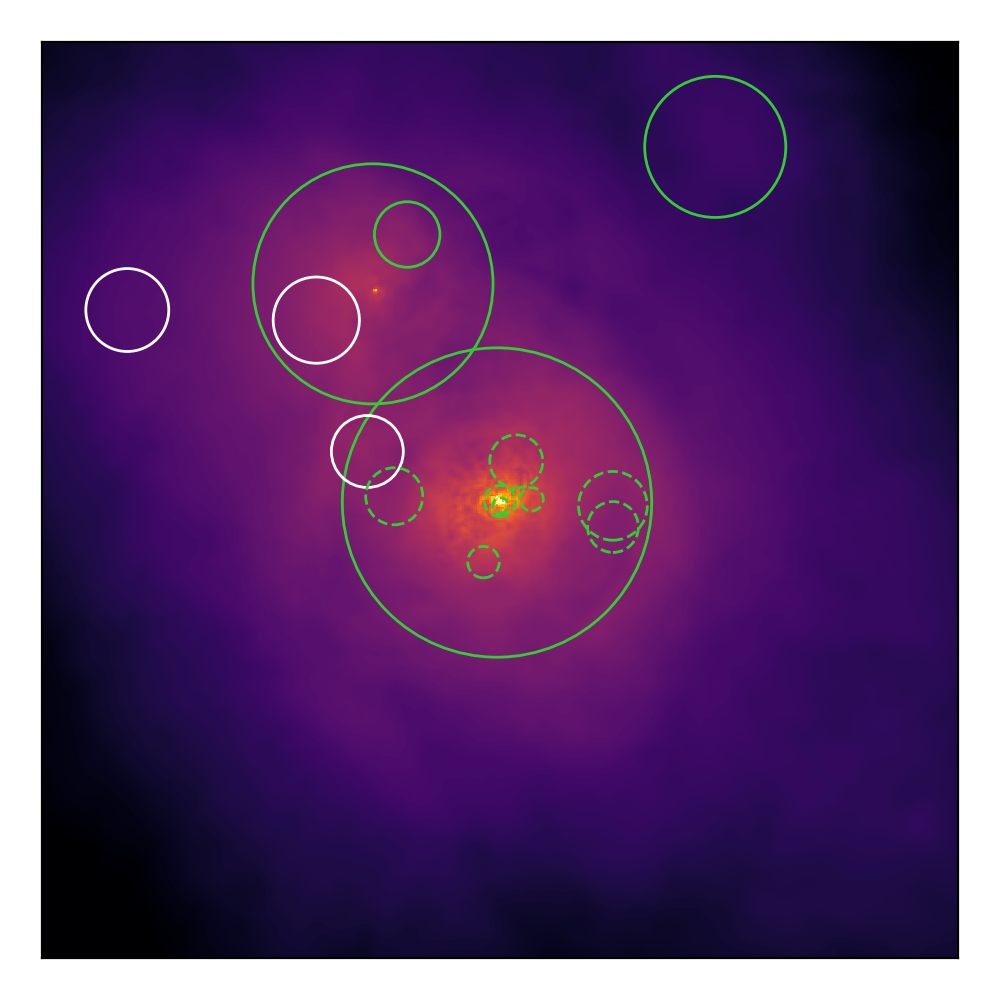}
    \caption{Projected DM (left) and gas (right) density in a cube of 500 kpc per side at $z=0$, centred on Bashful, the most massive halo in the simulation volume. The circles mark the viral radii of the identified galaxies, where the solid circles indicate the positions of the original seven dwarfs and dashed lines highlight the locations of the quenched satellite galaxies of Bashful. Green lines indicate halos that formed stars at some point during the simulation.}
    \label{fig:zoomregion}
\end{figure*}

We illustrate the selected zoom-region in Fig. \ref{fig:zoomregion}, where we show the projected DM density (left) and gas surface density (right) in a $\left(500\,\mathrm{kpc}\right)^3$ box centred on ``Bashful'', the most massive halo in the volume, at redshift zero. The solid circles indicate the locations and sizes of the seven most massive halos with non-zero gas content, with their radii corresponding to the virial radii. These galaxies are the same as the ones investigated in \citet{mina21} and \citet{shen_2014}. The four halos with star formation (Bashful, Doc, Dopey and Grumpy) are indicated with green lines, and the three starless halos are indicated with white lines. All galaxies are ``field'' dwarf galaxies, as the closest massive halo ($M_{vir} = 2.5 \times 10^{12} \rm M_{\odot}$) is located more than 3 Mpc away. Nevertheless, these galaxies themselves are interacting towards $z=0$, as shown clearly by the location of the circles. In addition, the simulation contains eight new satellite galaxies that appear to be completely quenched, which are labelled with the dashed circles. 

The global properties of the dwarf galaxies are shown in Table \ref{tab:dgs}. The seven named galaxies refer to the same halos as in \citet{shen_2014} and \citet{mina21}, and those labelled as \textit{qDG} refer to the population of quenched satellite galaxies of Bashful, which did not form stars in the previous runs. These galaxies have typical stellar mass around $10^{4}$ to $10^{5}$ solar masses, but only contain very old stellar populations, which suggests that they are relics from reionisation. We discuss them in detail in Section \ref{sec:qdgs}. In general, we found that the impact of radiation on galaxy properties strongly depends on the halo mass. Thus, we present the results for the two most luminous galaxies ($M_{vir} > 10^{10} \ \rm M_{\odot}$) first, followed by the fainter galaxies with $M_{vir} \sim 10^{9}-10^{10} \ \rm M_{\odot}$, and finally, we discuss the faintest population. 

\begin{table*}
    \caption{Properties of the simulated dwarf galaxies at redshift 0.}
    \centering
    \begin{tabular}{lcccccccccc}
    \hline \hline
Name & $M_\mathrm{vir}$ & $R_\mathrm{vir}$ & $V_\mathrm{max}$ & $M_\star$ & $M_\mathrm{gas}$ & $M_\mathrm{HI}$ & $f_\mathrm{b}$ & $\left<[\mathrm{Fe/H}]\right>$ & $M_\mathrm{V}$ & $B-V$\\
 & $[\mathrm{M_\odot}]$ & $[\mathrm{kpc}]$ & $[\mathrm{km\,s^{-1}}]$ & $[\mathrm{M_\odot}]$ & $[\mathrm{M_\odot}]$ & $[\mathrm{M_\odot}]$ & & & &\\
    \hline
Bashful & $3.47\times10^{10}$ & $84.43$ & $48.13$ & $4.96\times10^{7}$ & $9.79\times10^{8}$ & $3.49\times10^{7}$ & $0.030$ & $-1.40\pm0.94$ & $-14.54$ & $0.54$\\
Doc & $1.62\times10^{10}$ & $65.54$ & $36.98$ & $1.01\times10^{7}$ & $2.67\times10^{8}$ & $6.74\times10^{5}$ & $0.017$ & $-1.64\pm1.27$ & $-12.28$ & $0.67$\\
Dopey & $3.29\times10^{9}$ & $38.51$ & $22.42$ & $2.95\times10^{5}$ & $2.04\times10^{7}$ & $3.79\times10^{2}$ & $0.006$ & $-2.83\pm1.24$ & $-8.32$ & $0.69$\\
Grumpy & $1.77\times10^{9}$ & $17.88$ & $21.77$ & $7.51\times10^{5}$ & $6.37\times10^{6}$ & $2.02\times10^{2}$ & $0.004$ & $-2.74\pm1.24$ & $-9.38$ & $0.70$\\
Happy & $7.51\times10^{8}$ & $23.54$ & $15.54$ & --- & $8.67\times10^{5}$ & $3.39\times10^{1}$ & $0.001$ & --- & --- & ---\\
Sleepy & $4.36\times10^{8}$ & $19.63$ & $13.71$ & --- & --- & --- & --- & --- & --- & ---\\
Sneezy & $6.68\times10^{8}$ & $22.63$ & $14.90$ & --- & $2.79\times10^{6}$ & $7.35\times10^{1}$ & $0.004$ & --- & --- & ---\\
    \hline
qDG1 & $3.81\times10^{8}$ & $18.77$ & $13.37$ & $1.71\times10^{5}$ & $9.80\times10^{6}$ & $3.63\times10^{2}$ & $0.026$ & $-3.14\pm0.70$ & $-7.77$ & $0.70$\\
qDG2 & $2.42\times10^{8}$ & $14.47$ & $16.04$ & $6.33\times10^{5}$ & $4.45\times10^{4}$ & $4.69\times10^{0}$ & $0.003$ & $-2.37\pm0.78$ & $-9.15$ & $0.69$\\
qDG3 & $2.17\times10^{8}$ & $15.56$ & $13.07$ & $7.90\times10^{5}$ & $1.45\times10^{6}$ & $4.66\times10^{1}$ & $0.010$ & $-2.36\pm1.33$ & $-9.42$ & $0.70$\\
qDG4 & $1.54\times10^{8}$ & $13.88$ & $11.60$ & $1.18\times10^{5}$ & --- & --- & $0.001$ & $-2.58\pm0.92$ & $-7.32$ & $0.69$\\
qDG5 & $3.55\times10^{7}$ & $8.52$ & $9.34$ & $1.14\times10^{5}$ & --- & --- & $0.003$ & $-3.28\pm0.60$ & $-7.37$ & $0.71$\\
qDG6 & $3.55\times10^{7}$ & $8.58$ & $8.31$ & $8.26\times10^{4}$ & --- & --- & $0.002$ & $-3.23\pm0.45$ & $-6.99$ & $0.70$\\
qDG7 & $4.99\times10^{6}$ & $6.32$ & $3.64$ & $7.43\times10^{4}$ & --- & --- & $0.015$ & $-2.82\pm0.95$ & $-6.84$ & $0.69$\\
qDG8 & $2.10\times10^{6}$ & $4.23$ & $3.40$ & $2.48\times10^{4}$ & --- & --- & $0.012$ & $-2.66\pm1.03$ & $-5.61$ & $0.69$\\
    \hline
    \end{tabular}
    \label{tab:dgs}
    \tablefoot{Column 1 lists the name of the dwarf galaxy. Columns 2 to 11 provide the present-day viral mass, viral radius, maximum circular velocity, stellar mass, gas mass, HI mass, baryon fraction $f_\mathrm{b}=(M_\star+M_\mathrm{gas})/M_\mathrm{vir}$, mean stellar metallicity, $V$-band magnitudes and $B-V$ colour.}
\end{table*}

\subsection{Two Luminous Dwarfs}
\begin{figure*}
    \centering
    \includegraphics[width=\columnwidth]
    {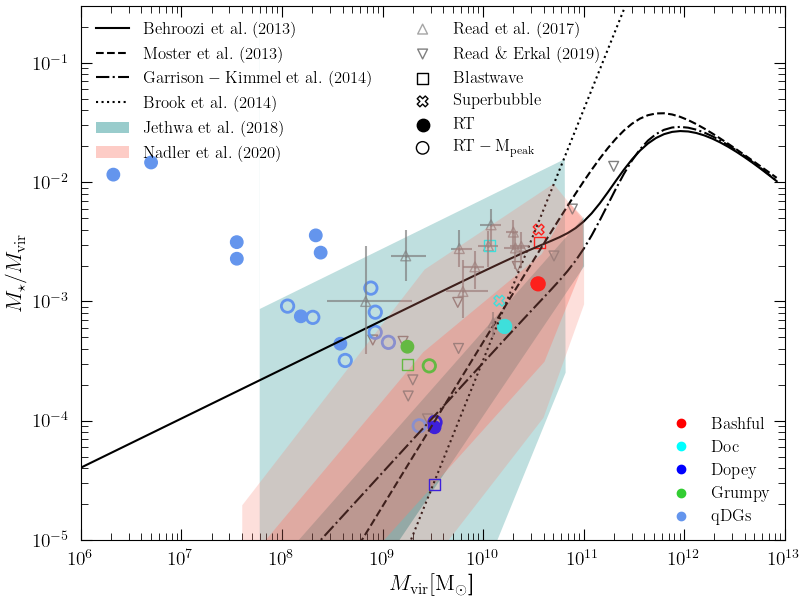}
    \includegraphics[width=\columnwidth]{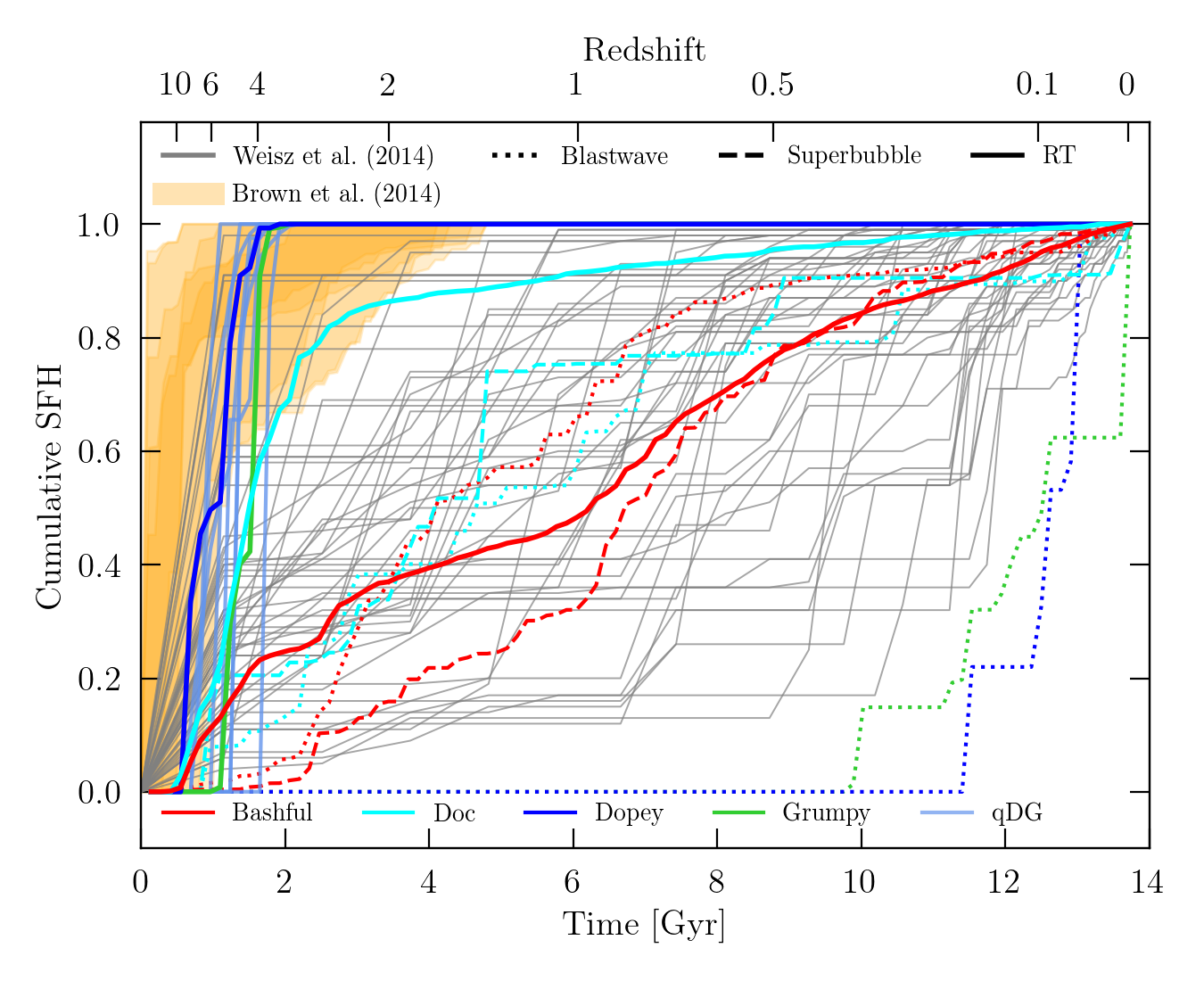}
    \caption{{\it Left panel:} Stellar mass-halo mass relation. The circles show the results of this work, where the filled ones show the stellar fraction as function of halo mass at $z=0$, and the open ones show results at the redshift where the halo mass is at its peak value (i.e. before tidal stripping during the accretion process). For comparison, the open crosses shows the \SB run from \citet{mina21} and the squares indicates the \BW run from \citet{shen_2014}. Open gray triangles represent the sample of isolated dwarf galaxies from \citet{read17} and satellite galaxies from \citet{read19}. The black lines show the mean present-day SMHM relation of \citet{behroozi13} (solid), \citet{moster13} (dashed), \citet{garrison-kimmel14} (dot-dashed) and \citet{brook14} (dotted). The blue and pink contours indicate SMHM relations from \citet{Jethwa18} and \citet{Nadler20}, respectively. The darker contour corresponds to 1$\sigma$ scatter and the lighter one indicates 2$\sigma$ scatter. {\it Right panel:} Cumulative star formation history of the simulated dwarf galaxies. Coloured lines represent the galaxies formed in the simulations. Solid lines show the results of the \RT run, colour coded as indicated in the Figure. Dotted and dashed lines correspond to the \BW and \SB runs, respectively. 
    Grey lines show the SFH derived form \citet{weisz14} for Local Group dwarfs and orange shaded regions show the SFH of ultra-faint dwarf galaxies from \citet{brown14}.}
    \label{fig:smhm}
\end{figure*}

\begin{figure*}
    \centering
    \includegraphics[width=\textwidth]{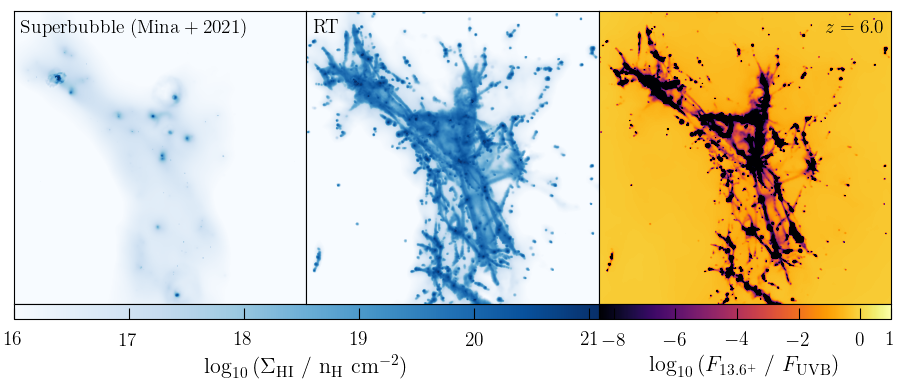}
    \caption{Projected HI density ($\Sigma_\mathrm{HI}$) in a cube with 1 comoving Mpc per side at $z=6$ for the \SB (left) and \RT (middle) run. Right: Density weighted radiation field at $13.6^+$ eV relative to the \citet{HM12} UVB flux.}
    \label{fig:SigmaHI}
\end{figure*}

The stellar mass-halo mass (SMHM) relation and the star formation history of the simulated galaxies are shown in Figure \ref{fig:smhm}. The present day stellar mass ($M_{\star}$) of the two most massive dwarf galaxies with $M_{vir} > 10^{10} \rm \ M_{\odot}$, Bashful and Doc, are $5.0 \times 10^{7} \ \rm M_{\odot}$ and $1.0 \times 10^{7} \ \rm M_{\odot}$, which correspond to a stellar mass fraction of 0.0014 and $6.2 \times 10^{-4}$, respectively. These are in excellent agreement with the observational data of isolated and satellite dwarf galaxies from the Local Group \citep{read17,read19} and the extrapolation of SMHM relations from abundance matching (AM) technique \citep{moster13, brook14, garrison-kimmel14, Jethwa18, Nadler20} although somewhat below the AM relation derived by \citet{behroozi13}. The comparison between this work (filled circles) and the run with superbubble feedback only (open crosses) from \citet{mina21} indicates that the radiative feedback suppresses star formation by a factor of 2-3 in this classical dwarf galaxy regime. 

The cumulative star formation history shows that both Bashful and Doc are star-forming at $z=0$. Comparing the RT simulation results (solid line) with superbubble feedback only (dashed line), we find that the inclusion of RT increases the fraction of old stellar populations. In particular, Doc has more than 80\% of stars formed at $z\ge3$. This is mostly due to the increase in early star formation before and during reionisation. Unlike the typical approach of turning on the uniform UV background instantaneously as used in \citet{shen_2014} and \cite{mina21}, with RT, the UV background radiation is emitted from virtual background sources, distributed close to the edge of the simulation box. As the ionisation front associated with the UVB has to propagate through the IGM, it does not reach the halos immediately. This time delay\footnote{We note that this time delay is due to the R-type expansion of the ionisation front, rather than the speed of light. {\sc TREVR2} assumes infinite speed of light as it is an instantaneous method.}  prevents gas from heating up and allows it to cool and collapse to higher densities, enabling them to form stars during the EoR. This effect is demonstrated in Figure \ref{fig:SigmaHI}. The left and middle  panels show the neutral hydrogen column density $\Sigma_{\rm HI}$ at $z = 6.0$ from the \citet{mina21} superbubble simulation, and our simulation with RT, respectively. With RT there is significantly more HI gas at this redshift, especially in the halos and the filaments connecting them where the radiation has not yet penetrated through. This is shown in the right panel of Figure \ref{fig:SigmaHI}. The flux of ionising radiation within halos and filaments is much smaller than the UV background value. Focusing back on the SFH properties in Figure \ref{fig:smhm}, we find that the impact of radiation on the star formation history is dependent on galaxy halo mass, with lower mass halos (Doc) being more sensitive to radiation and photo-heating than higher mass halos (Bashful), and this effect is even more obvious for the smallest mass halos (cf. Section\ref{faintdwarfs}).
\begin{table*}[]
    \centering
    \caption{Bashfuls quiescent satellite population.}
    \begin{tabular}{ccccccccc}
    \hline\hline
        Name & $z_\mathrm{SF,first}$ & $z_\mathrm{SF,last}$ & $\Delta t_\mathrm{SF}$ [Myr] & $z_\mathrm{infall}$ & $M_\mathrm{HI,infall}$ [$\mathrm{M_\odot}$] & $M_\mathrm{peak}$ [$\mathrm{M_\odot}$] & $M_\mathrm{vir}/M_\mathrm{peak}$  & $\sigma$ [km/s] \\
    \hline
        qDG1 & 4.857 & 4.141 & 275.5  & 0.59 & 135.1            & 2.32$\times10^9$ & 0.16   & 8.67 \\
        qDG2 & 8.098 & 3.172 & 1460.3 & 1.70 & 86.6             & 8.47$\times10^8$ & 0.29   & 7.64 \\
        qDG3 & 5.469 & 3.985 & 525.0  & 1.39 & 127.6            & 7.65$\times10^8$ & 0.28   & 6.56 \\
        qDG4 & 7.030 & 4.165 & 746.9  & 1.18 & 15.3             & 4.26$\times10^8$ & 0.36   & 7.78 \\
        qDG5 & 3.881 & 3.575 & 170.0  & 2.43 & 97.9             & 2.03$\times10^8$ & 0.17   & 6.41 \\
        qDG6 & 4.784 & 3.806 & 415.3  & 1.75 & 0.89             & 1.14$\times10^8$ & 0.31   & 5.79 \\
        qDG7 & 5.992 & 4.596 & 388.5  & 2.66 & $2.7\times10^3$  & 1.14$\times10^9$ & 0.0044 & 2.73 \\
        qDG8 & 6.953 & 5.833 & 206.8  & 3.35 & $3.8\times10^3$  & 8.45$\times10^8$ & 0.0025 & 2.00 \\
    \hline
    \end{tabular}
     \tablefoot{Column 1 lists the name of the dwarf galaxy. Columns 2 and 3 are the redshift of the first and last star formation event. Column 4 gives the time interval from the first to last star formation event in Myr. Columns 5 and 6 are the redshift when the halo approaches $2\times R_\mathrm{vir}$ of Bashful and the HI mass at this redshift. Columns 7 and 8 are the peak halo mass and the ratio of the present day virial mass to peak halo mass. Column 9 lists the line-of-sight velocity dispersion.}
    \label{tab:qDGs}
\end{table*}

\subsection{Two fainter dwarfs}
\label{faintdwarfs}
Dopey and Grumpy are the two fainter galaxies with halo mass $M_{vir} \sim 10^{9}-10^{10} \ \rm M_{\odot}$ and stellar mass fraction of 9.0 $\times 10^{-5}$ and 4.2 $\times 10^{-4}$, respectively. Although this is again in agreement with the abundance matching results, it is interesting that these two galaxies do not form in the SN feedback-only simulation with the superbubble model \citep{mina21}. As shown in the right panel of Figure \ref{fig:smhm}, in the RT simulation the stellar population is old, with the majority of star formation occurring before $z \sim 4$. More precisely, Dopey and Grumpy start forming their stars during the era of reionisation ($z=8.52$ and $5.5$ respectively) and cease SF by $z=3.47$ and $3.27$, respectively. Such an SFH is very similar to the observed Ultra-Faint Dwarf galaxies in the Local Group \citep[e.g.,][]{brown14}, although Dopey and Grumpy have somewhat higher masses than typical UFDs \citep[$M_{v} > -7.0$,][]{simon19}. The early star formation is again related to the propagation of the UVB in the RT simulation, which allows the IGM to be mostly neutral and cold for a longer time, so that it can accrete onto small halos like the progenitors of Dopey and Grumpy. No star formation occurred at lower redshift when the reionisation was complete, which is consistent with the \SB run results. Nevertheless, the same halos form galaxies in the \BW run in \citet{shen_2014} with completely opposite star formation histories (dotted lines in the right panel of Figure \ref{fig:smhm}). They have similar mass but predominantly young stellar populations. Our results highlight how sensitive the formation of the faintest dwarf galaxies depends on reionisation history, radiative feedback, SN feedback models, and the chemical and thermodynamic states of the circumgalactic gas.

\subsection{The quenched satellites} \label{sec:qdgs}

Comparing to the \BW and the \SB runs, we find eight more quenched dwarf galaxies (qDGs) in the simulation volume, all of which are satellites of Bashful at $z = 0$. These galaxies have halo masses ranging from $10^{6}$ M$_{\odot}$ to a few times $10^{8}$ M$_{\odot}$ at $z = 0$, but with typical stellar mass $\sim 10^{4}-10^{6}$ M$_{\odot}$. The three smallest galaxies qDG6, qDG7 and qDG8 have $M_{v} > -7.0$ and can be classified as UFDs. The smallest halos have stellar mass fractions much higher than the extrapolation of the stellar mass-halo mass relations. However, as shown in Table \ref{tab:qDGs}, the halo masses before accretion ($M_{\rm peak}$) are significantly larger, ranging from $10^{8}$ to a few times $10^{9}$ M$_{\odot}$, indicating strong tidal stripping of dark matter during the accretion process. The ratios between current and peak halo mass for these halos are less than 40\%, and can be as low as 0.25-0.44\% (qDG8 and qDG7, respectively). The stellar mass fraction as a function of $M_{vir, peak}$ is fully consistent with the SMHM relations from \citet{behroozi13} and \citet{Jethwa18} and observations of isolated galaxies from \citet{read17}. Similar to Dopey and Grumpy, the qDGs contain only old stellar populations, are red in B-V colours, and are in agreement with the SFHs of observed UFDs. It is worth noting that the halos "Happy", "Sleepy", and "Sneezy" all have masses above $10^{8}$ M$_{\odot}$, which is similar to or higher than the peak halo mass of some qDGs, and yet there is no early star formation.

\begin{figure}
\centering
    \includegraphics[width=\columnwidth]{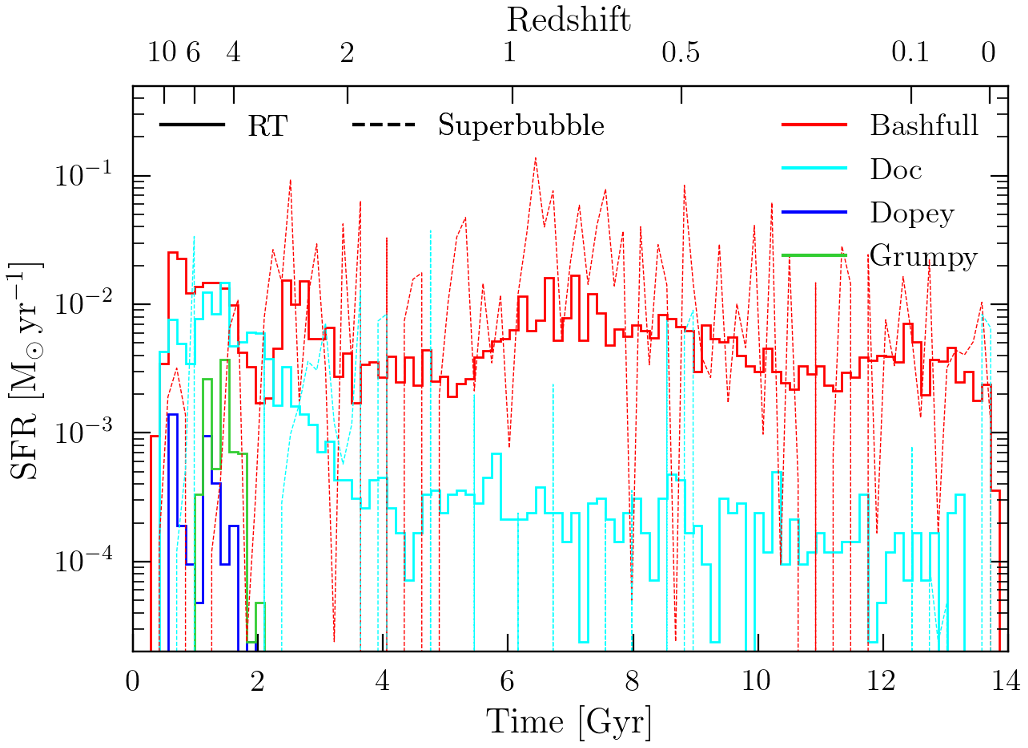}
    \caption{The star formation rate of Bashful, Doc, Dopey and Grumpy as a function of time and redshift. The thin, dashed lines with similar colours indicate the results from the \citet{mina21} run which is without RT but with the same superbubble feedback.} 
    \label{fig:sfr}
\end{figure}

\begin{figure}
    \centering
    \includegraphics[width=\columnwidth]{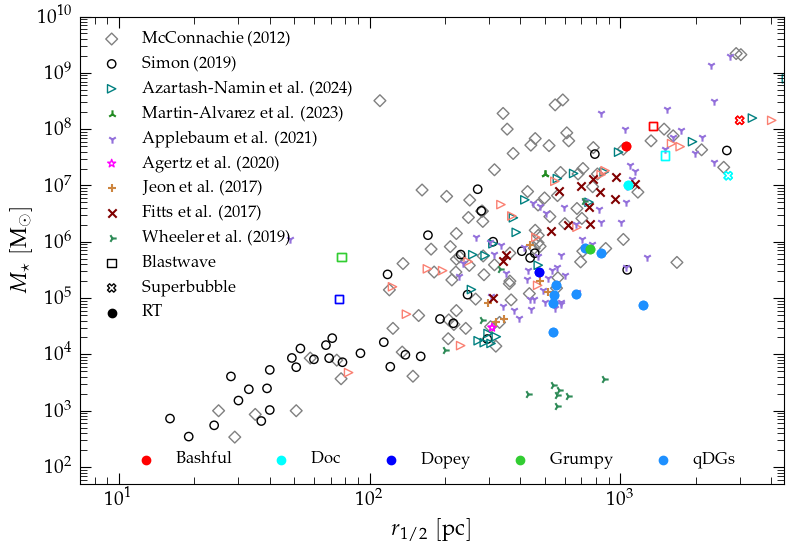}
    \caption{Stellar mass versus half-light radius of the simulated dwarf galaxies, compared to observations form \citet{mcconnachie12} and \citet{simon19} and simulations form \citet{jeon17,fitts17,wheeler19,applebaum21,Azartash-Namin24} and RT simulations from \citet{agertz20} and \citet{martin-alvarez23}. For the \citet{Azartash-Namin24} data points, the teal and light red colour corresponds to their \BW and \SB run, respectively.
    }
    \label{fig:MassSize}
\end{figure}

\subsection{Star formation rate}
\label{sec:sfr}
In Figure \ref{fig:sfr}, we show the star formation rates of Bashful, Doc, Dopey and Grumpy. The \SB run from \citet{mina21} is also plotted for comparison (where only the two most massive halos form galaxies). When RT is included, although the star formation histories are still bursty with rapid changes of the SFR, the amplitude of the changes is generally much smaller than the ones in the \SB run. 
It is likely that radiative feedback and photoheating precondition the ISM, such that the star formation occurs in smaller clumps, which results in weaker collective SN feedback and galactic outflows and hence less suppression of star formation. This is consistent with previous RT simulations of idealised disc galaxies \citep[e.g.,][]{Rosdahl2015}. Bursty star formation caused by disruptive outflows is often associated with the formation of dark matter cores, as these outflows often result in a rapid change of the central gravitational potential wells \citep[e.g,][]{Governato2010, Pontzen2012}. As radiative feedback renders the star formation less bursty, it is expected that the formation of the dark matter core is also impacted. We discuss this in detail in Section \ref{sec:dm}. 

As star formation is suppressed less strongly by SNe, the SFH in both Bashful and Doc lacks an extended quiescent period, during which SF is completely quenched. This is different from the \SB simulation, especially for the relatively lower mass Doc. Observationally, resolved stellar population analysis for the Local Group dwarf galaxies does show breaks of star formation activity for tens to hundreds of million years or longer in some galaxies \citep{Tolstoy09}, which may indicate the collective effect of feedback in these systems is stronger than in our simulation. However, the observed SFHs are still highly diverse among different galaxies and long quenched periods are not common \citep[e.g.,][]{McQuinn24}. In general, a quiescent period does not necessarily imply that feedback is more "ejective". Preventive processes such as photoionisation heating can quench dwarf galaxies (as it does for Dopey, Grumpy, and all the qDGs in our simulation). In addition, environmental factors such as mergers and ram-pressure stripping cannot be neglected for satellite dwarfs. As our simulation sample is small, it is not sufficient to address the diversity of SFHs in various mass halos and environments. Theoretically, many other simulations also demonstrate a strong burstiness of their SFH and long quiescent times \citep[e.g.,][]{Maccio17, fitts17, Wright19, applebaum21}, but again, there is no clear observational evidence that observed dwarf galaxies should be as bursty \citep{Governato2015}. It is recently shown by \citet{Zhang24} that such burstiness is sensitive to the implementations of SN feedback, in particular the directions of the momentum injection. Our result further indicates that the SFH is sensitive to the interplay between radiation and SN feedback.  

\begin{figure}
    \centering
    \includegraphics[width=\columnwidth]{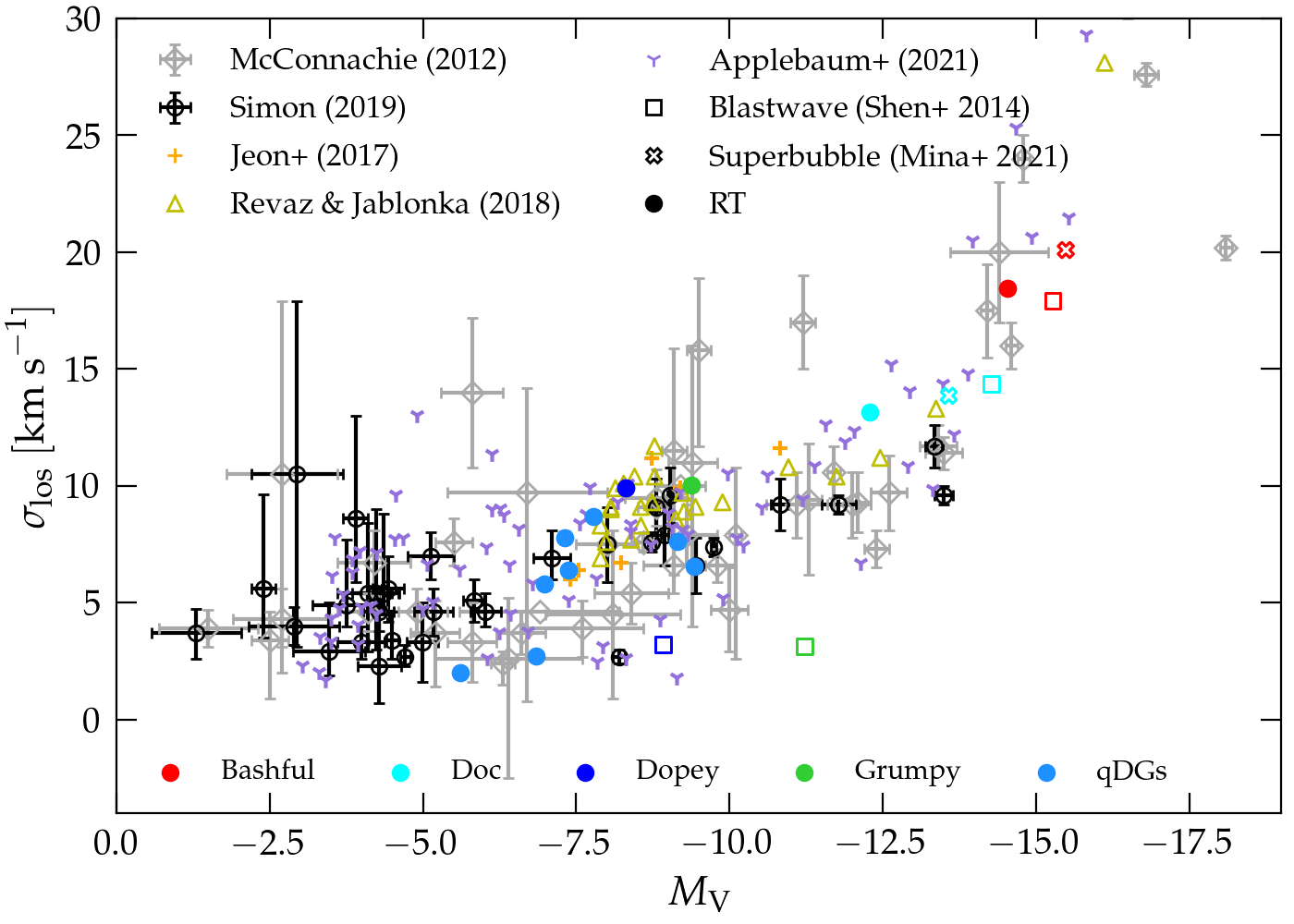}
    \caption{$M_\mathrm{V}-\sigma_\mathrm{los}$ relation of the simulated dwarf galaxies compared to observations form \citet{mcconnachie12} and \citet{simon19} and simulations from \citet{jeon17}, \citet{revaz18} and \citet{applebaum21}. 
     Solid circles show the results of the new \RT run, colour coded as indicated in the Figure, open squares and crosses correspond to the \BW and \SB runs. 
    }
    \label{fig:MvSigma}
\end{figure}

\subsection{Stellar distribution and kinematics}
In Figure \ref{fig:MassSize}, we compare the stellar half-light radius, $r_{1/2}$, of our simulated galaxies to the previous Seven Dwarf simulations, various simulations from other groups, and also observations from \citet{mcconnachie12} and \citet{simon19}. For simplicity and comparison with other simulations, our $r_{1/2}$ is the circular half-light radius which is directly computed from the simulation data by summing up luminosity from each stellar particle from the center until half of the total luminosity is reached. We note that this gives slightly different results from the observationally estimated $r_{1/2}$.The simulated galaxies in this work (solid dots) have $r_{1/2}$ between 476 pc and about 1.2 kpc,
which are generally consistent with observations and other simulation works. Comparing to the \SB run (empty crosses in the figure), we note that the inclusion of radiation reduces the stellar half-light radius significantly for Bashful and Doc, by about a factor of 2-3, which makes it more consistent with observations. This is an interesting result, and again indicates that feedback-driven outflows are weaker with RT. Early galaxy formation simulations with ineffective feedback often result in compact, bulgy galaxies \citep[e.g.,][]{Scannapieco12}, which is known as the "angular momentum catastrophe". Galactic outflows are shown to eject lower angular momentum gas and thus enable disc galaxies to form \citep[e.g.,][]{Brook2011}. Moreover, \citet{mina21} compared the superbubble feedback with the delayed-cooling blastwave feedback (shown also here in figure \ref{fig:MassSize} in open squares), and concluded that the superbubble model drives stronger outflows with higher mass-loading factor than in the delayed-cooling model, and thus creates more extended stellar discs. However, it is worth noting that the behaviour of a feedback model depends on star formation models and total energy input. In \citet{Azartash-Namin24}, the authors compared the superbubble and blastwave feedback models (triangle symbols with light red and teal colours, respectively) in the Storm simulations, but found no clear difference in galaxy sizes between the two. It is likely because the Storm simulation uses an H$_{2}$-based star formation model which alters the burstyness of star formation, and a higher energy input per SNe than the \BW run. In the current work, the reduction of $r_{1/2}$ clearly suggests that outflows are less strong with RT, possibly as a result of less-clustered star formation, as discussed in Section \ref{sec:sfr}. 

As expected, the smaller galaxies Dopey, Grumpy, and the qDGs usually have smaller sizes than Bashful and Doc. It is worth noting that the drastically different SFHs in Dopey and Grumpy between our simulation and the ones in \citet{shen_2014} also lead to different galaxy sizes. Dopey and Grumpy in \citet{shen_2014} are blue and very compact, with half-light radii only 80-90 pc (dark blue and green open squares), whereas in our simulation they are much more extended ($r_{1/2}$ about 476 pc and 759 pc for Dopey and Grumpy, respectively). \citet{shen_2014} found that these two galaxies have formation histories and morphologies similar to extremely metal-deficient blue compact dwarfs \citep[XBCD;][]{Papaderos2008}, and argued that the compactness is due to the dynamically unsettled stellar population. In our simulation, these two dwarfs are similar to normal UFDs with only an ancient stellar population, and in this case they reach dynamical equilibrium with the gravitational potential wells.

In Figure \ref{fig:MvSigma}, we show the relation between the V-Band magnitude and the 1D line-of-sight velocity dispersion. The simulated galaxies have a $M_{v} - \sigma_{los}$ relation in good agreement with observations of Local Group dwarfs from \citet{mcconnachie12} and UFDs from \citet{simon19}: the velocity dispersion decreases with decreasing luminosity (mass) and plateaus at $M_{v} \gtrsim -10$, although the scatter is high towards the faint end. Comparing with the \SB (open crosses) and \BW (open squares) simulations, we find that the inclusion of RT or varying feedback implementations does not strongly affect the stellar velocity dispersion for Bashful and Doc, which indicates that stellar kinematics are less sensitive to radiative and/or mechanical feedback in the ISM. Similar conclusions are also reached in \citet{agertz20} with the EDGE dwarf simulations with various resolutions, feedback, and also RT. Nevertheless, for Dopey and Grumpy, our simulation produced a much higher velocity dispersion than that of the \BW run, possibly also due to the fact that the SFH is drastically different between the two simulations, and the galaxies in the \BW run have entirely young stellar populations and thus are not dynamically settled. 

Although internal processes such as feedback and RT may not strongly impact stellar kinematics, dynamical interactions and tidal stripping are likely to play an important role. All our quenched satellites (qDGs) have smaller velocity dispersion compared to the field dwarfs, even for the ones with similar or higher stellar mass (e.g., qDG2, qDG3 compared to Dopey and Grumpy see Table \ref{tab:dgs}). As noted in Section \ref{sec:qdgs}, these halos experience severe tidal stripping when they accrete onto Bashful. As demonstrated most recently in \citet{applebaum21} with the Justice League simulation of dwarf satellites in the Local Group,  $\sigma_{los}$ of satellite galaxies appear to be proportional to $M_{vir, z = 0}/M_{peak}$ (which indicates the severity of stripping), and in their simulations, the majority of galaxies with low velocity dispersion ($\sigma < 5$ km/s) are stripped when accreting onto the Milky Way. Our simulation confirms this result. In fact, the only two galaxies with $\sigma < 5$ km/s in our sample are qDG7 and qDG8, which have the strongest stripping and the smallest $M_{vir}$ at $z = 0$ (cf. Table \ref{tab:qDGs}).  Tidal interactions reduce stellar velocity dispersion, even though the host halo (Bashful) is relatively small compared to the MW-mass halos in \citet{applebaum21}.

\begin{figure}
    \centering
    \includegraphics[width=\columnwidth]{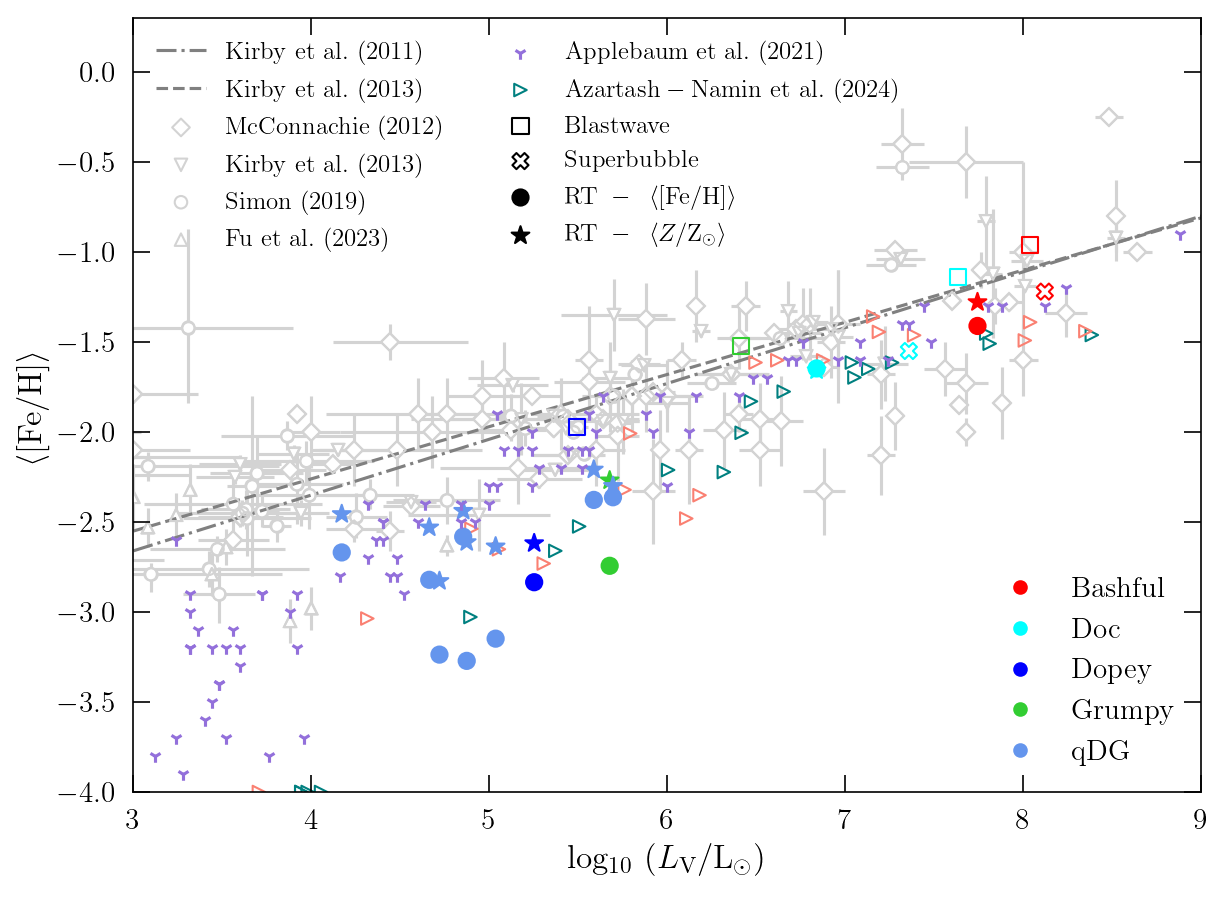}
    \caption{Stellar luminosity-metallicity relation of the simulated dwarf galaxies. The solid circles show the results of this work, and the coloured crosses and squares indicate the previous Seven Dwarfs simulation of \citet{mina21} (superbubble feedback, no RT) and \citet{shen_2014} (blastwave feedback, no RT), respectively. The grey symbols represent observations of Milky Way and Andromeda satellite galaxies \citep{mcconnachie12,kirby13}, open black circles and open stars show observations of local UFDs from \citet{simon19} and \citet{Fu23}. The gray lines show the best fit to the LZR of the Milky Way dwarf satellite galaxies \citep{kirby11} and of Local Group dwarf galaxies \citep{kirby13}. For comparison, we also plot simulation data from \citet{applebaum21} (purple symbols) and \citet{Azartash-Namin24} (teal and light red triangles for their \BW and \SB simulations, respectively). } 
   \label{fig:massmetal}
\end{figure}

\begin{figure}
    \centering

    \includegraphics[width=\columnwidth]{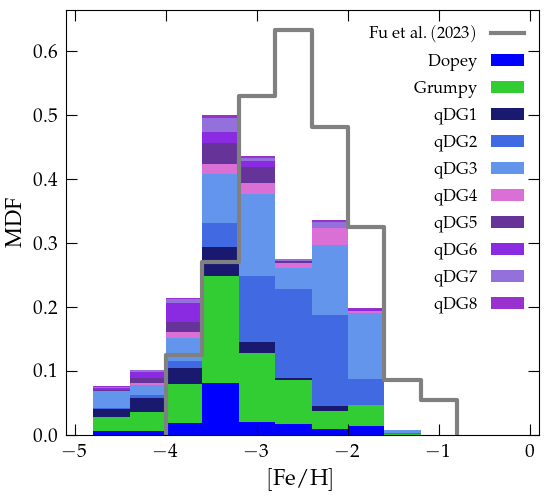}
    \caption{The metallicity distribution function for Dopey, Grumpy and all qDGs (all galaxies with $L_{v} < 10^{6}$ L$_\odot$). To compare with the UFD observations from \citet{Fu23}, we have removed stellar particles with [Fe/H] < -4.8 from the sample. The composite MDF of 13 Milky Way UFDs from \citet{Fu23} is plotted with a gray line.}
    \label{fig:MDF1}
\end{figure}

\subsection{Metallicities} 
\label{sec:metallicity}

The mass-metallicity relation (MZR) or luminosity-metallicity relation (LZR) is one of the fundamental scaling relations observed in galaxies across many orders of magnitude in mass. For dwarf galaxies in the Local Group (both dSphs and dIrrs), it has been shown from resolved stellar population studies that they follow a universal MZR(LZR) despite different photometries and morphologies \citep[e.g.][]{kirby11, kirby13}. Dwarf galaxies are typically metal-poor, and chemical evolution models suggest that a large fraction (if not the majority) of metals are ejected from galaxies. The MZR extends well into the UFD mass regime, although the metallicities appear to plateau around [Fe/H] $\sim -3.0$ at the faintest end, and the relation has significantly larger scatter \citep{simon19}. Figure \ref{fig:massmetal} shows the stellar LZR compared to the observations of the Local Group satellite and irregular dwarf galaxies \citep{mcconnachie12,kirby13} and the UFDs of \citet{simon19} and \citet{Fu23}. When computing the average metallicity, here we do not add a metallicity floor of [Fe/H]=-4.0, as has been done in some previous work \citep[e.g.][]{fitts17,applebaum21} to account for the early enrichment from Population III stars. But in practice, adding such a floor does not impact our results. 

Similarly to previous Seven Dwarf simulations, the metallicities of Bashful and Doc are fully consistent with the observed MZR. Compared to the \SB run (cross symbols), our RT simulation decreases the average stellar metallicity proportionally to the decrement in stellar mass, such that the data points still have a similar slope as the \citet{kirby11} fitting function. This is particularly interesting for Doc, as it has a much older stellar population than the previous \SB simulations (cf. Figure \ref{fig:smhm}, right panel), and yet the metallicity does not decrease significantly, which suggests that the high-z ISM of Doc retains more metals. 

For fainter galaxies with $L < 10^{6}\, \mathrm{L_\odot}$, the average metallicity is systematically lower than the observations, especially for the satellites qDG1, qDG5, qDG6 (which have <[Fe/H]> < -3.0) and the dIrrs Dopey and Grumpy. To understand this, we show the metallicity distribution function (MDF) for all faint galaxies in our simulation in Figure \ref{fig:MDF1}. We compare with recent observations of 13 MW UFDs from HST narrowband imaging by \citet{Fu23}. Following the procedure in \citet{Fu23}, we exclude stellar particles with [Fe/H]< -4.8. Note that some galaxies in our sample have slightly higher stellar mass, so they are not strictly UFDs, but such a comparison can still be illustrative. In general, the simulation overproduces the number of low-metallicity stars with [Fe/H] < -3.0, while lacking stars with higher metallicities. The peak of the MDF is around [Fe/H] $\sim$ -3.5 in our simulated faint galaxies, but around -2.5 for observed UFDs. The fraction of stellar particles with [Fe/H] < -3.0 for qDG1, qDG5 is qDG6 88\%, 85\% and 89\%, respectively. It is therefore unsurprising that the average metallicities are below the observed MZR for these galaxies.  Several factors can contribute to this. First, feedback processes may eject too much metal from the ISM. For example, \citet{agertz20} showed that boosting SN feedback energy significantly decreases [Fe/H] in the EDGE simulations. It was argued that MZR or LZR is a critical test to differentiate galaxy formation models in simulations. Although ejective feedback is possibly the most important factor in reproducing the MZR, a second factor is the relative abundances of different elements. For stellar LZR, iron abundances are most often measured to represent metallicities. However, iron yields are still highly uncertain, especially from Type Ia supernovae. More recent stellar evolution and yield models for SN II \citep[e.g.][]{Pignatari18} and SN Ia \citep[e.g.][]{Leung18} significantly boost Fe production at an early time \citep[e.g.][]{Hopkins23}. In Figure \ref{fig:massmetal}, we show the actual average metallicity ($Z/Z_{\odot}$) of the simulated galaxies (star symbols). Here, the simulation appears to be more consistent with the observations, especially at the fainter end. It is possible that our yield models adopted from \citet{raiteri96} underestimate iron abundances, as they are relatively older models. Similar results are also found in the Justice League simulation in \citet{applebaum21}, which uses the same yield models as in our work. We defer to future work to investigate the role of chemical enrichment models in the mass-metallicity relations. One of the strong, but complementary constraints to understand the MZR is the metal content in the circumgalactic and the intergalactic media, but CGM/IGM observations are not without challenges, especially for faint, satellite dwarf galaxies, where their CGM is often shared with the more massive host. We discuss the CGM from our simulation in Section \ref{sec:cgm}. 

In Figure \ref{fig:massmetal}, we also show data from the Justice League simulation from \citet{applebaum21} and the Storm simulation from \citet{Azartash-Namin24} for comparison, since the three simulations use similar hydrodynamic codes, metal diffusion, and yield models. The main differences are that both \citet{applebaum21} and \citet{Azartash-Namin24} use an H$_{2}$-based star formation model and they do not have RT. The Justice League simulation uses the blastwave feedback model of \citet{stinson06} and simulates Local Group analogues, whereas the Storm suite uses both the blastwave and the superbubble feedback models and simulates more isolated dwarf systems (although the most massive halo is still larger than our simulation, with M$_{vir} \sim 10^{11}$ M$_{\odot}$). It is interesting that the faint galaxies within $L_{V} \sim 10^{4} - 10^{6}$ L$_{\odot}$ in \citet{applebaum21} generally have higher metallicities than our simulation and are more consistent with observations (although still at the lower end), but the \citet{Azartash-Namin24} results are more similar to ours. A possible reason is pre-enrichment from the nearby massive systems in the Justice League simulation as it simulates a LG-like environment. Another possibility is differences in the quenching time. The UFDs in \citet{applebaum21} appear to have more extended SFHs compared to ours (cf. Figure 11 in their article). Although mostly quenched by $z \sim 3$, there is a significant fraction of galaxies in their simulation with star formation lasting $> 1$Gyr. In contrast, our qDGs are mostly quenched completely before 500 Myr, with the exception of qDG2, which has a SFH extended to 1.5 Gyrs. qDG2 has an average metallicity of [Fe/H] = -2.37, which is more consistent with observations. However, even the \citet{applebaum21} simulation still seems to quench the SF too rapidly compared to the UFD observations, in particular from \citet{weisz14}.

\section{Formation of Dark Matter Cores}\label{sec:dm}

\begin{figure}
    \centering
    \includegraphics[width=\columnwidth]{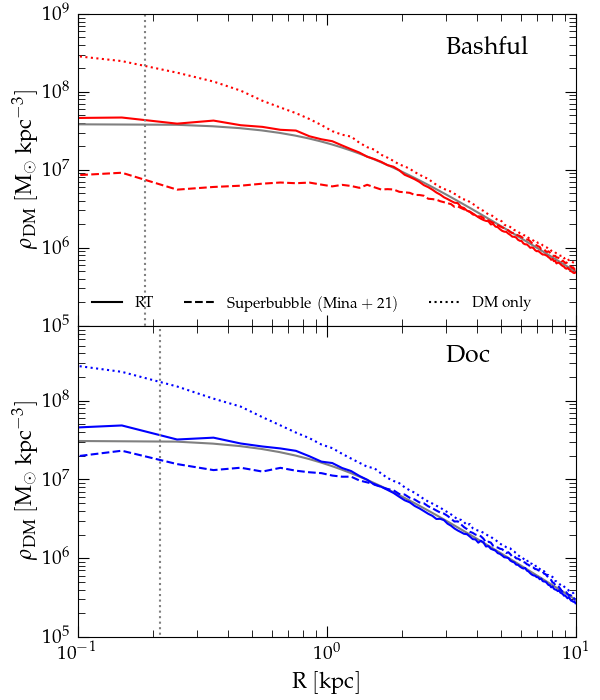}
    \caption{Present-day dark matter density profile for the two most massive dwarf galaxies in the simulation, Bashful (top) and Doc (bottom). The solid lines represent the RT run, the dashed lines correspond to the \SB run \citep{mina21}, and the dotted line is the DM only comparison run. The gray line shows the best fitting pseudo-isothermal DM profile. The vertical dashed lines indicate the radii below which numerical convergence can not be reached to due two body relaxation \citep{power03}.}
    \label{fig:DMprofile}
\end{figure}

\begin{figure}
    \centering
    \includegraphics[width=\columnwidth]{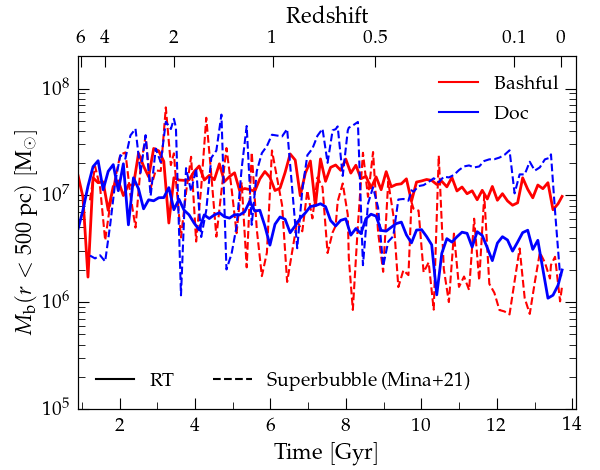}
    \caption{Evolution of baryonic mass within the inner 500 pc of Bashful (red) and Doc (blue). Solid lines represent the RT run and dashed lines show the \SB run \citep{mina21}.}
    \label{fig:Mbary}
\end{figure}

Recent works have shown how stellar feedback, in particular from repeated supernova explosions, can transform the inner dark matter profiles of dwarf galaxies, thereby the central cusp is transformed into a core \citep[e.g.][]{readgilmore05,Governato2010,governato12,penarrubia12,Pontzen2012,madau14,onorbe15,fitts17,mina21}.
We expect the changes in the SFH and thus available feedback energy between the \SB run from \citet{mina21} and the \RT run to alter the properties of the dark matter cores in the simulated dwarf galaxies. In Fig. \ref{fig:DMprofile}, we show the DM density profiles of Bashful (top) and Doc (bottom) from our simulation. For comparison, the results of the \SB simulation from \citet{mina21} and a dark-matter only simulation are also plotted. It is clear that core formation still occurs in our simulation, but the core sizes are significantly smaller than in the \SB simulation. 
As in the \SB and \BW runs, the dark matter distribution is well described by a pseudo-isothermal density profile, defined as:
\begin{equation}
    \rho_\mathrm{DM}=\frac{\rho_0}{1+\left(r/r_\mathrm{c}\right)},
\end{equation}
where $\rho_0$ is the central DM density and $r_\mathrm{c}$ is the core radius. The best fitting pseudo-isothermal density profiles are also shown in Figure \ref{fig:DMprofile} (gray lines). At $z=0$ the DM core sizes of Bashful and Doc are $r_c=1.117\pm0.014$ kpc and $0.969\pm0.014$ kpc, respectively, which are about 3 (2) times smaller than those of the \SB runs for Bashful (Doc). In other words, the addition of radiative feedback decreases DM core sizes. The average central dark-matter densities within 150 pc, $\rho_{c}(150 \ \rm pc)$, for Bashful and Doc are $5.76 \times 10^{7} \rm M_{\odot} \ \rm kpc^{-3}$ and $4.69 \times 10^{7} \rm M_{\odot} \ \rm kpc^{-3}$, respectively. This is in agreement with the data from \citet{read19b}, where the authors found that the central DM densities within 150 pc for local dwarf galaxies are below $10^{8} \ \rm M_{\odot} \ \rm kpc^{-3}$ if the galaxies have more extended star formation. 

The reason for the reduced core sizes is likely due to the fact that Bashful and Doc have less bursty star formation in our simulation. As discussed in Section \ref{sec:sfr}, a smoother SFR indicates that star formation is less clustered and the SN feedback is weaker. As a result, the fluctuations of central baryonic densities in our simulation are not as prominent as in previous SN-only runs. We illustrate this in Fig. \ref{fig:Mbary}, where we show the evolution of baryonic mass within the central 500 pc of Bashful and Doc for the \RT and \SB runs. In the \RT run, Bashful's central baryonic mass remains around $10^7~\mathrm{M_\odot}$ with variations of only about a factor of two. In contrast, Bashful in the \SB run show much larger fluctuations of its central mass, often exceeding an order of magnitude. A similar difference between the two runs is also seen in Doc.  In the \SB simulation (as well as in the previous \BW simulation, not shown in Figure. \ref{fig:Mbary}), the strong burst of SF nearly evacuates the gas in the central regions of these halos, followed by a quiescent phase during which the gas reservoir is replenished before the next SF episode repeats the cycle. Such cycles cause rapid and strong fluctuations of the central potential wells and eventually result in dark matter particles irreversibly moving to larger distances, and thus form a core \citep{Pontzen2012}. In the RT simulation these cycles still occur, but with smaller amplitudes, which leads to a smaller DM core. This is consistent with \citet{Azartash-Namin24}, where the authors compared a large number of dwarf halos within the Storm simulation and found clear positive correlation between the burstiness of SF and DM core slopes. Back to Figure \ref{fig:Mbary}, it is interesting to note that, unlike Bashful, Doc's central density in our simulation is on average lower than in the \SB simulation, despite that SN feedback is weaker. This suggests radiative feedback may have played a role in preventing gas accretion and reaccretion after SN ejections, and this impact is more significant in lower-mass galaxies. 

\begin{figure}
    \centering
    \includegraphics[width=\columnwidth]{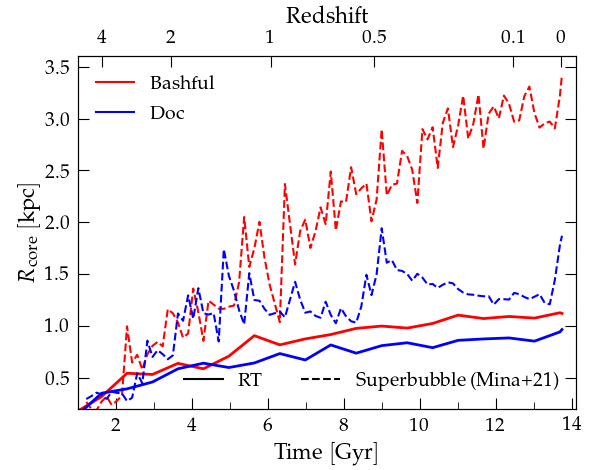}
    \caption{Temporal evolution of the DM core radius for Bashful (red) and Doc (blue). Solid lines represent the \RT run and dashed lines show the \SB run \citep{mina21}. The less bursty SFH and overall smaller stellar mass in the \RT run results in significantly smaller DM cores. }
    \label{fig:Rcore}
\end{figure}

The milder baryonic mass fluctuations at the centre of the halo appear to correlate with the core size evolution, which is shown in Figure \ref{fig:Rcore}. Unlike the \SB run (dashed lines) and previous \BW runs \citep[Figure 3 in][]{madau14}, the evolution of core sizes for both Bashful and Doc in our simulation is more monotonic: core formation starts at an early time, as in previous simulations, but there are no abrupt increases and decreases in core sizes within a short period of time. In previous SN-only simulations, a strong gas evacuation period is usually followed by a sudden increase in the size of the core, while rapid gas reaccretion causes the size of the core to decrease significantly due to adiabatic contraction, in which the DM halo compresses and becomes more cuspy when a large amount of baryonic matter is accreted at the centre \citep[e.g.][]{Sellwood05}. With the amplitudes of both gas ejection and accretion reduced, our simulated galaxies have a smoother core size evolution. Nevertheless, in all simulations the dark-matter core is persistent and survives gas inflows from reaccreting previous outflows, and from merger events.

As in \citet{mina21}, we estimate the minimum energy required for the cusp-core transformation and compare it to the total energy injected by Type II SNe ($\Delta E_\mathrm{SN}$). The energy required for the cusp-core transformation can be estimated as $\Delta W/2 \equiv (W_\mathrm{core} - W_\mathrm{cusp})/2$ \citep{penarrubia12}, where $W$ is the gravitational binding energy of the DM halo, given by:
\begin{equation}
    W = -4\pi\, G \int_0^{R_\mathrm{vir}}\rho_\mathrm{DM}\,M_\mathrm{enc}\,R\,\mathrm{d}R,
\end{equation}
where $G$ is the gravitational constant, $\rho_\mathrm{DM}$ is the DM density and $M_\mathrm{enc}$ is the total mass enclosed within radius $R$. Transformation from a cuspy to a cored DM profile is possible if $\Delta E_\mathrm{SN}$ is greater than $\Delta W/2$.
At $z=0$, $(\Delta E_\mathrm{SN},\,\Delta W/2) = (9.05\times10^{56}\,\mathrm{erg},\,5.04\times10^{56}\,\mathrm{erg})$ for Bashful and $(1.87\times10^{56}\,\mathrm{erg},\,9.09\times10^{55}\,\mathrm{erg})$ for Doc. For both dwarfs, the available energy from SN explosions is still abundant, about twice that of the minimum energy needed for cusp-core transformation. The difference in DM mass between the hydrodynamic run and the DM-only run within the core radius, $\Delta M_{DM} \equiv M_{DM}^{cusp} - M_{DM}^{core}$, is $1.41\times10^8 \mathrm{M}_\odot$ and $1.00\times10^8\,\mathrm{M}_\odot$ for Bashful and Doc, respectively. The ratio between $\Delta M_{DM}$ and the stellar mass $M_{\star}$ roughly indicates how effective stellar feedback kinematically heats dark matter. This "DM core removal efficiency" ($\Delta M_{DM}/M_{star}$), is 2.8 for Bashful and 9.9 for Doc. It seems that stellar feedback is more efficient to create cores in lower mass halos, until $M_{\star}$ become too small so that feedback is not sufficiently energetic \citep{penarrubia12,read19b}. However, our RT simulation appears to have a much smaller $\Delta M_{DM}/M_{star}$ compared to the SN-feedback only \SB run, which has $\Delta M_{DM}/M_{star} $ of 6.55 (19.39) for Bashful (Doc). Again, radiative feedback modulates the baryonic cycles and renders core formation less efficient.

\begin{figure}
    \centering
    \includegraphics[width=\columnwidth]{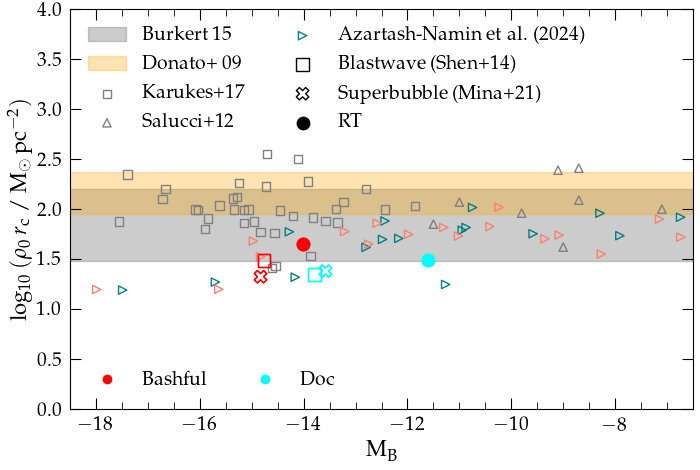}
    \caption{DM core surface density ($\rho_0\,r_\mathrm{c}$) as a function of B-band magnitude of Bashful and Doc compared to disk dwarf galaxies \citep[][gray squares]{karukes17}, Milky Way dSphs \citep[][gray triangles]{salucci12} and the scaling relations from \citet[yellow shaded region]{donato09}, \citet[gray shaded region]{burkert15} and simulation data from \citet{Azartash-Namin24} (teal and light red triangles for their blastwave and superbubble feedback simulations, respectively.
    }
    \label{fig:rho0rc}
\end{figure}

The product of central DM halo density and the DM core radius, $\rho_0\, r_\mathrm{c}$, has been found to be remarkably constant over a wide range of galaxy types and a luminosity range of more than 14 magnitudes in the B band \citep[e.g.][and references therein]{salucci19}.
For example, \citet{donato09} found $\mathrm{log}\,(\rho_0 r_c)=2.15\pm0.2$ based on co-added rotation curves of spiral galaxies, mass modelling of individual (dwarf) galaxies, and galaxy-galaxy weak-lensing signals, with indications that the relation holds true for dSphs. Subsequent works found that the product $\rho_0\, r_\mathrm{c}$ of the Milky Way's dSphs is consistent with extrapolations from higher-mass systems \citep[e.g.][]{salucci12,burkert15}.
In Figure \ref{fig:rho0rc}, we compare $\rho_0\, r_\mathrm{c}$ of Bashful and Doc to the scaling relations from \citet{burkert15} and \citet{donato09} and to disk dwarf galaxies from \citet{karukes17} and the dSph from \citet{salucci12}. In addition, we compare results from \citet{Azartash-Namin24} of the Storm Simulations. The higher central DM density and smaller core size of Bashful and Doc in the \RT run make $\rho_0\, r_\mathrm{c}$ more consistent with the derived scaling relations of galaxies, while in both the \SB and \BW Seven Dwarf simulations this product is somewhat too low. However, there is considerable scatter in the $\rho_0\, r_\mathrm{c}$-M$_{B}$ relation in both observational data and simulations with a larger galaxy sample such as the Storm simulation, indicating other factors, such as environments and merger histories may play a role. Future RT simulations with a larger sample of dwarf galaxies in different environments are necessary to further understand the impact of radiative feedback on cusp-core transformation. 

Our results are in agreement with many previous works in the literature, which demonstrated that impulsive baryonic outflows and inflows triggered by supernova feedback can dynamically heat dark matter and form DM cores in dwarf galaxies \citep[e.g.][]{Mashchenko08, Governato2010, governato12, onorbe15,fitts17}. Adding radiative feedback in our simulation mainly reduces the strength of the baryonic cycle and thus decreases the DM core sizes. However, it is still under debate whether SN-feedback is sufficient for cusp-core transformation even for classical dwarfs. For example, no core formation was found in the APOSTLE simulations of the Local Group environment  \citep{Sawala16} and the AURIGA simulations of MW-like halos \citep{Bose19}, nor were DM cores found in isolated dwarf simulations from \citet{revaz18} of halos up to virial mass of $\sim 10^{10}\, \rm M_{\odot}$. More recent work (Pandora Project) from \citet{martin-alvarez23} simulated a $10^{10}\, \rm M_{\odot}$ halo with increasing complex physics, and concluded that SN feedback-only fails to form DM cores at $z \sim 3.5$, and radiative transfer and cosmic ray feedback contribute to the core formation, which is opposite to what we find here.

In halos with $\Delta E_{SN} > \Delta W/2$ such as Bashful and Doc, in principle, there is enough energy in SNe to "remove" the DM from the central region and lift it to a larger orbit \citep{penarrubia12}. However, in practice it is highly dependent on how much the SNe energy is coupled with the ISM, which regulates the episodic cycle of outflows and inflows, and the disturbance of the central gravitational potential well. Indeed, within the same framework of the Seven Dwarfs simulations with the same initial condition, same hydrodynamic and gravity solvers, and heating and cooling functions etc., variations of SN feedback model alone can alter the DM core size by 1-2 kpc. The superbubble model generates outflows with lower velocity and larger mass-loading factor than the previously adopted delayed cooling model \citep{keller14}, it creates larger fluctuations of the central mass density and thus larger DM cores. In general, although various subgrid SN feedback implementations can converge relatively well on reproducing global properties of galaxies, it has been shown that the outflow properties can look drastically different in both idealised simulations \citep[e.g.][]{Smith19} and cosmological comparison simulations such as AGORA \citep{Strawn24}. Given the strong correlation between outflows and the core formation, it is not unexpected that DM core formation and core sizes vary significantly. 

It is interesting to note that simulations with resolved Sedov-Taylor phase for SN feedback \citep{agertz20,Gutcke21,Deng24} have also not reached consensus on cusp-core transformation.  For the EDGE simulations, DM heating and flattening of the central profile are observed in the simulated halos, but the cusp-core transformation is not only caused by stellar feedback, but also by frequent minor mergers at later times, after the halos are quenched by reionisation at $z \sim 4$. However, late major mergers can regenerate the cusp \citep{Orkney21}. Thus, the core formation depends on both SF activities and mass assembly histories. In contrast, LYRA simulations do not produce DM cores \citep{Gutcke22}. We note that these are simulations of UFDs with halo mass $M_{vir} \sim 10^{9} \rm\, M_{\odot}$ and stellar mass typically less than $10^{5}-10^{6} \rm\, M_{\odot}$, and SN-feedback may not provide sufficient energy to be solely responsible for cusp-core transformation. Moreover, as they are often quenched at a high redshift, the impact of assembly histories (major and minor mergers) can be significant, resulting in large scatter in DM core properties at z = 0. Nevertheless, it is a promising approach, and it will be interesting to also apply it to more massive dwarf galaxy simulations. Observations of inner DM content in this mass regime also have large uncertainties because many galaxies do not have HI gas or clean HI kinematics, and as such, it is difficult to measure the rotation curves.  

For the fainter dwarf galaxies in our simulation, only Dopey, Grumpy, and qDG2 have marginally resolved DM cores ($r_\mathrm{c}=0.290$ kpc, $0.344$ kpc and $0.490$ kpc, respectively, after we apply the \citet{power03} criterion to estimate the radius within which the two-body relaxation time is shorter than the Hubble time. These three galaxies are among the most massive ones in the faint dwarf population and have stellar masses of several $10^{5}\rm\, M_{\odot}$, similar to the aforementioned work on UFD simulations. However, it is interesting that qDG3 does not have a resolved DM core, although it has a higher stellar mass than all the three galaxies that have cores. As qDG3 is quenched earlier ($z_{\rm SF, last}$ = 4.0, cf. Table \ref{tab:qDGs}), we speculate that even if a small core formed in this galaxy due to early SF activities, later merger events can transform it back to a cusp. This is in agreement with a recent study using the EDGE simulations by \citet{Muni25}, in which it was found that early SF quenching by reionisation is correlated with higher central DM density at $z=0$. However, due to the limited resolution in our simulation, we will not perform a further analysis of the faint dwarfs here and defer it to future work.

\section{The Circumgalactic Medium}\label{sec:cgm}

\begin{figure*}
    \centering
    \includegraphics[width=\textwidth]{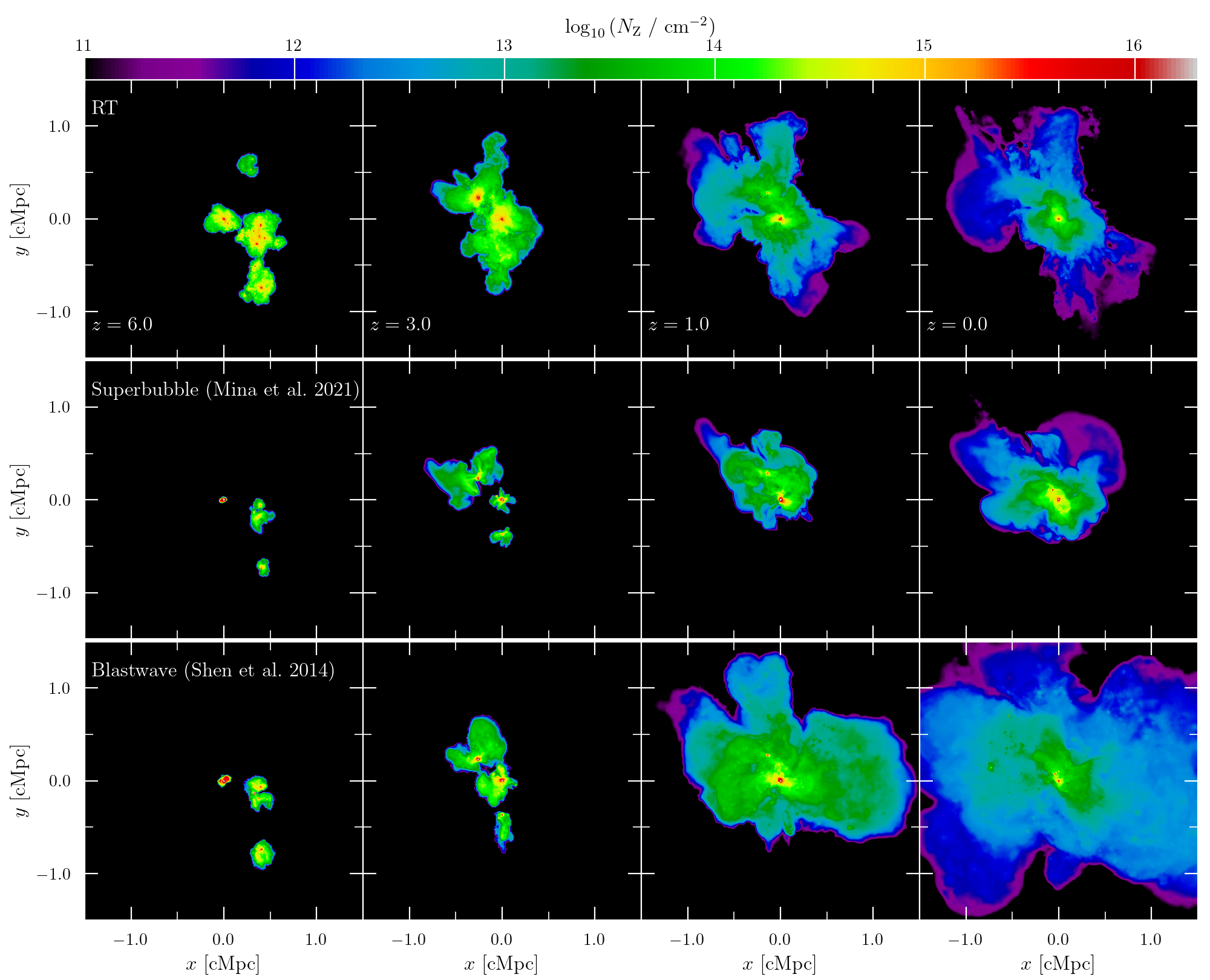}
    \caption{Metal distribution through cosmic time from z=6 to z=0 for the \RT (top), \SB (middle) and \BW (bottom) runs. Metals in the \RT run are distributed over a greater area as stars are formed in a greater number of small progenitor halos. 
    }
    \label{fig:MetalCGMTimeCompareSims}
\end{figure*}

\begin{figure*}
    \centering
    \includegraphics[width=\textwidth]{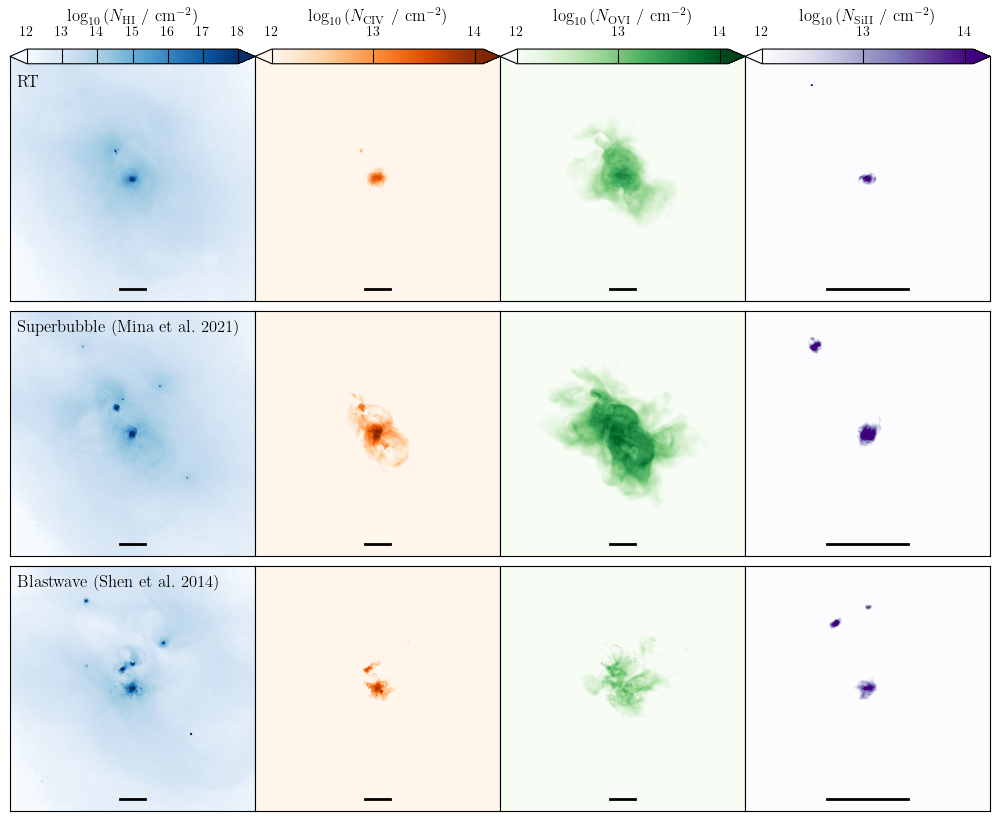}
    \caption{Column density of \ion{H}{I}, \ion{C}{IV}, \ion{O}{VI} and \ion{Si}{II} (from left to right) at $z=0$ for the \RT, \SB and \BW runs (top to bottom). The black line in each panel indicates a length of 100 kpc. 
    }
    \label{fig:MetalCGMcompare}
\end{figure*}

\begin{figure*}
    \centering
    \includegraphics[width=\columnwidth]{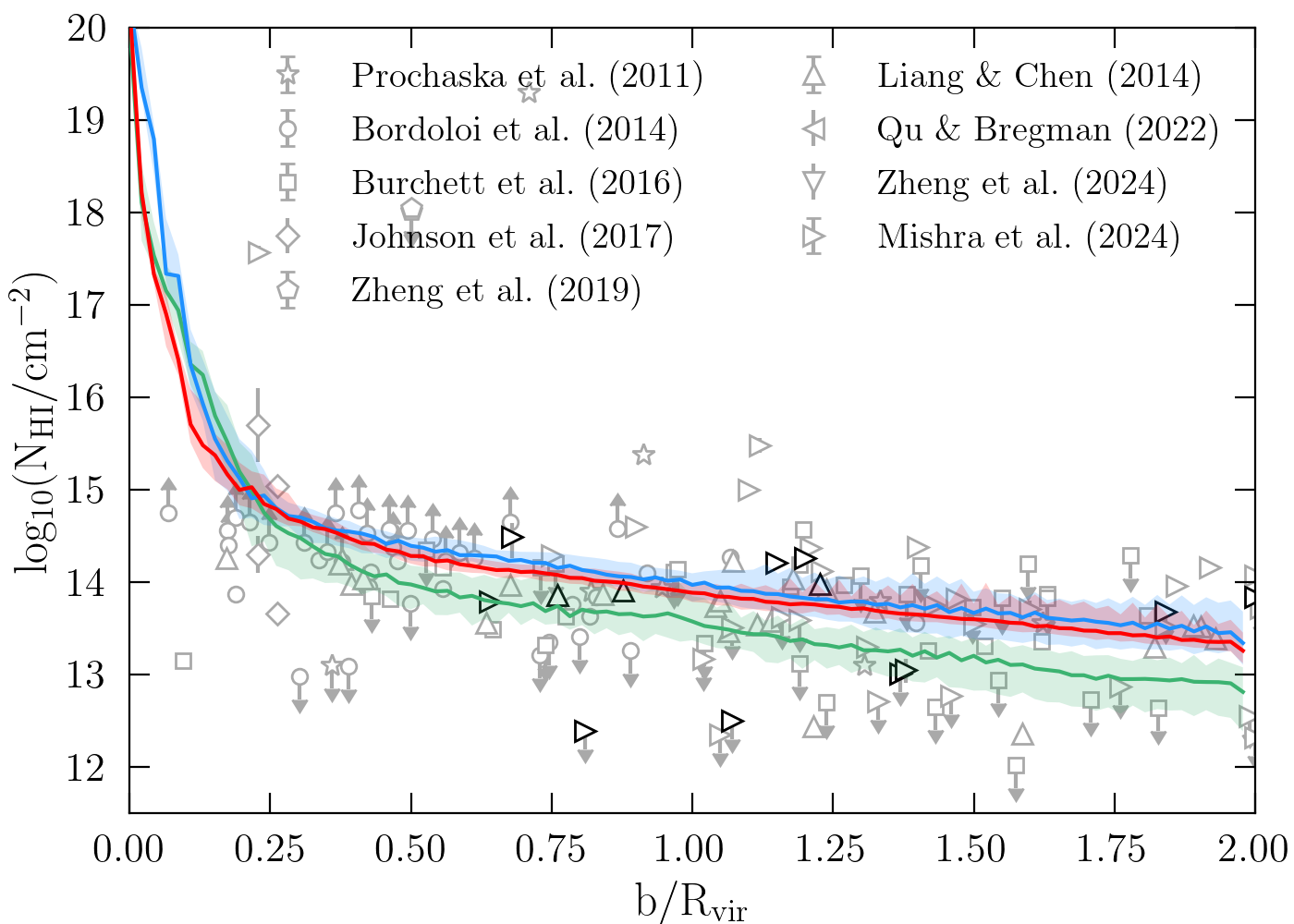}
    \hfill
    \includegraphics[width=\columnwidth]{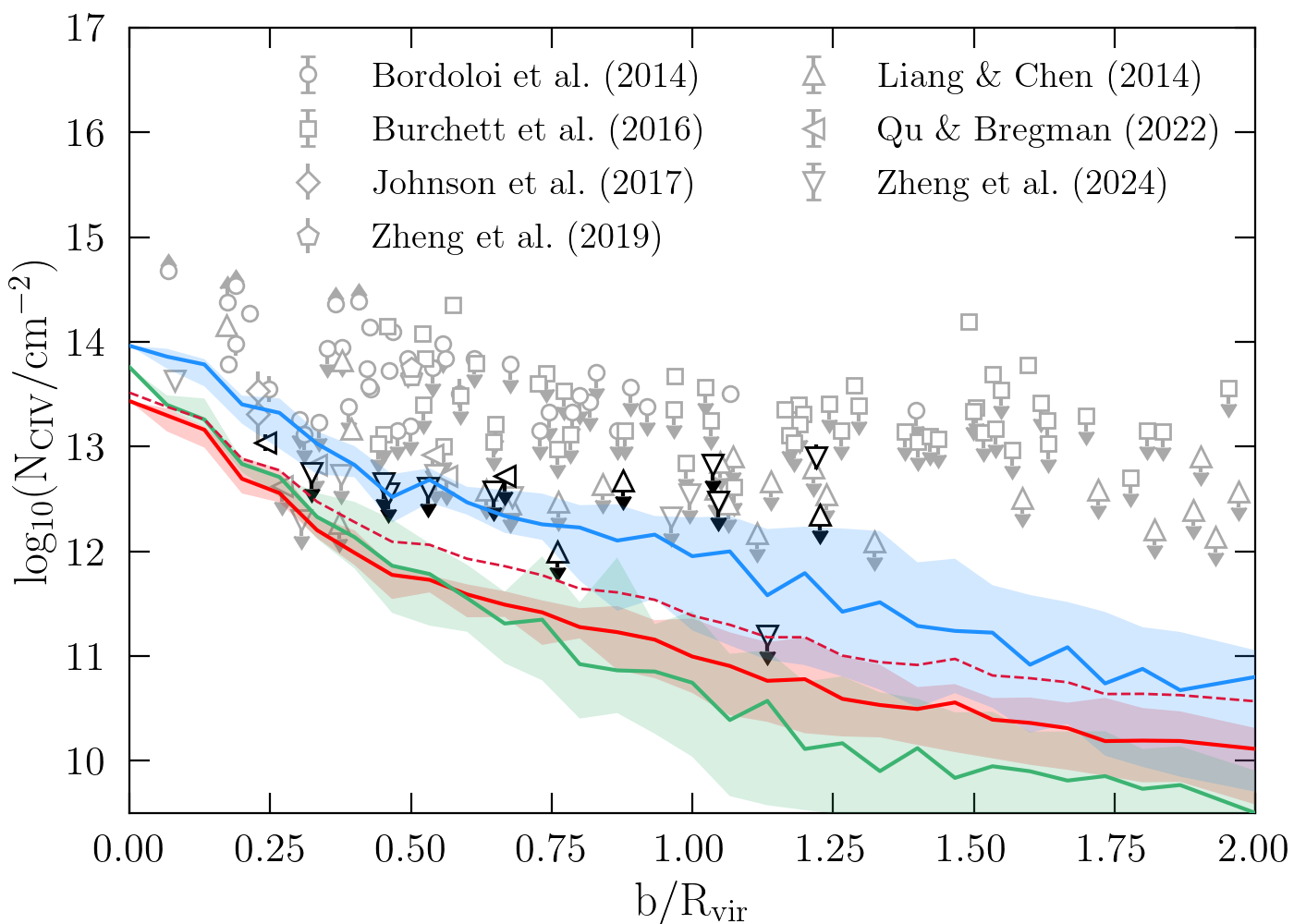} \\
    \includegraphics[width=\columnwidth]{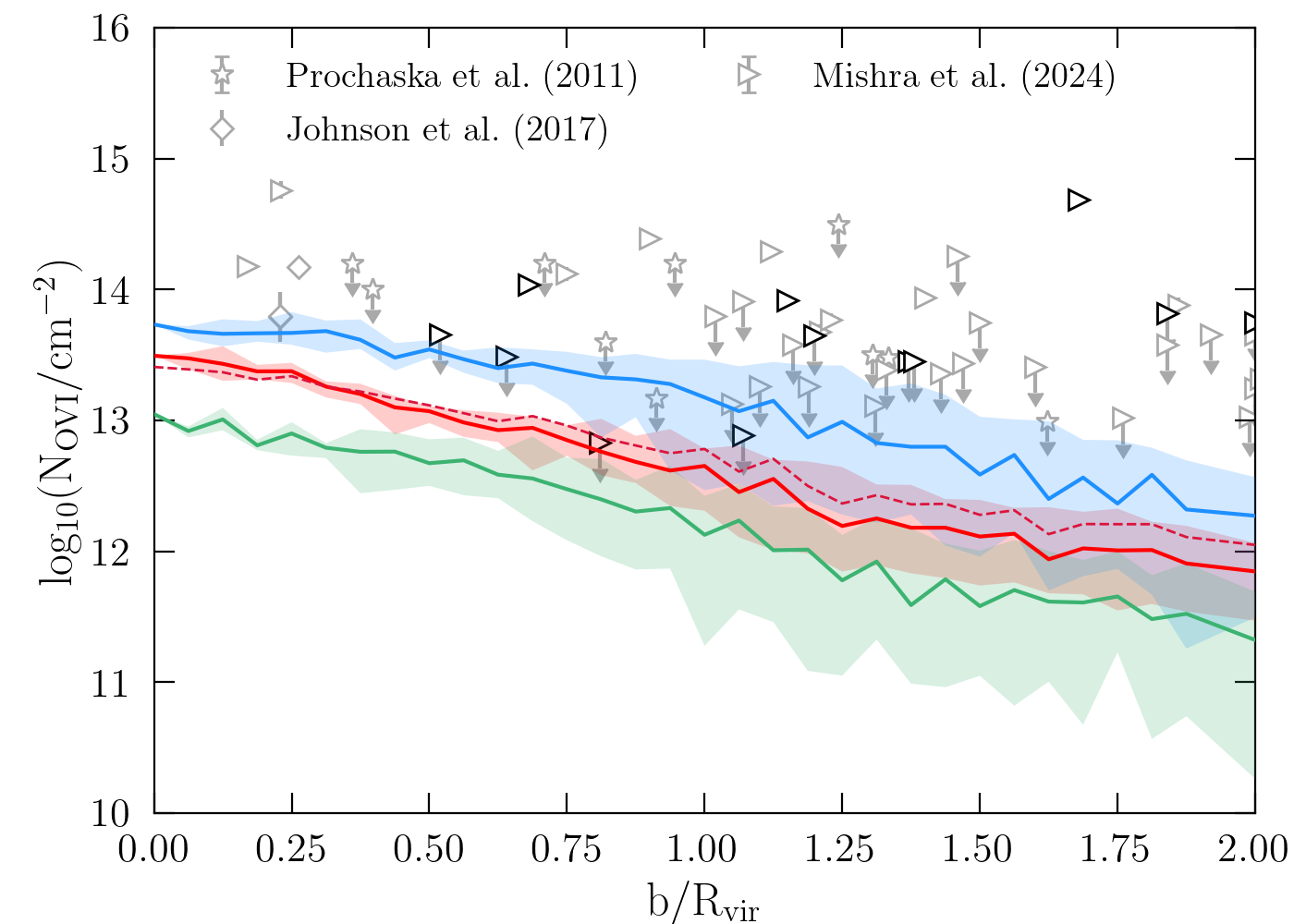}
    \hfill
    \includegraphics[width=\columnwidth]{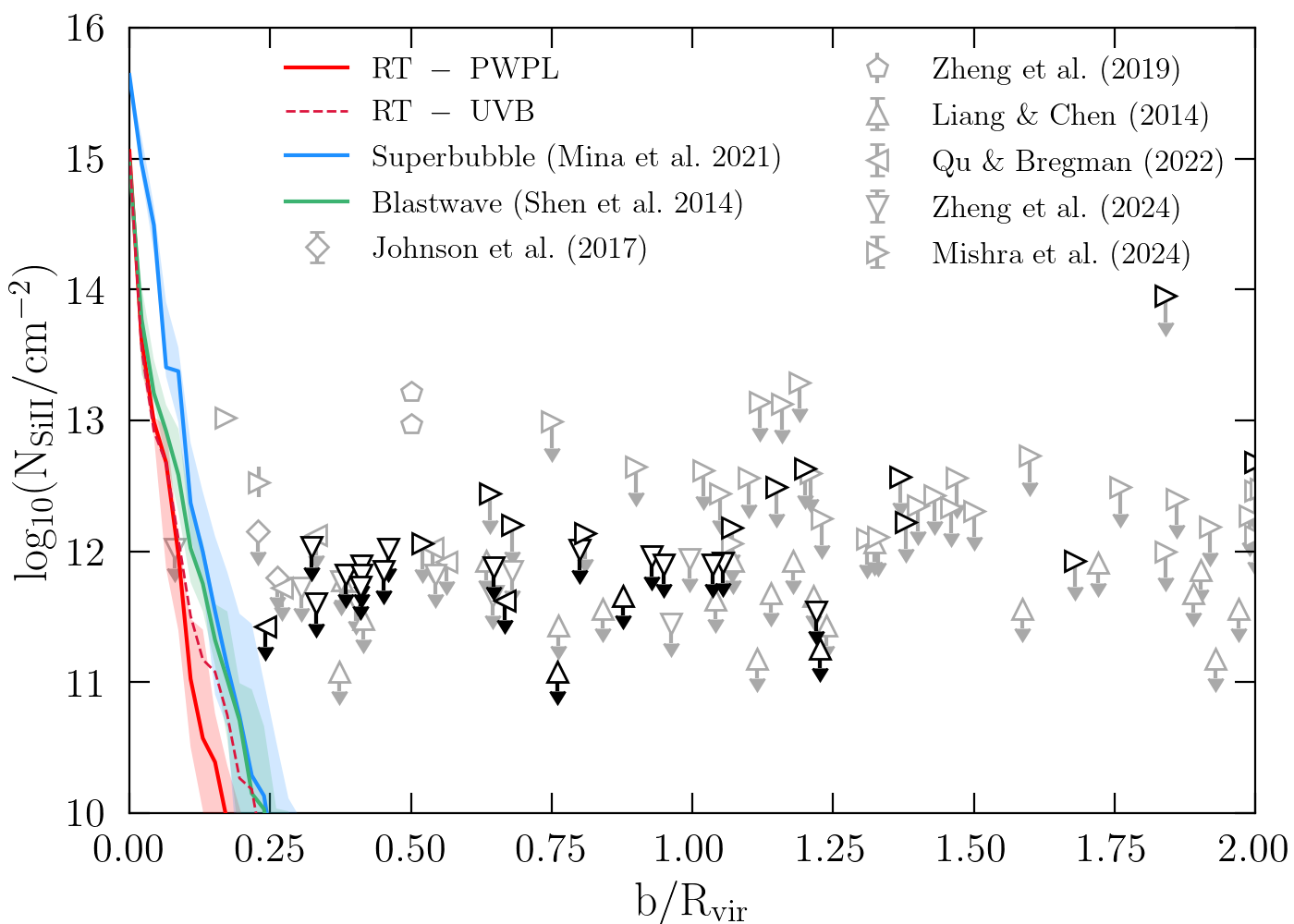}
    \caption{Present-day column density as a function of impact parameter scaled with the virial radius of Bashful ($b/R_{vir}$) of \ion{H}{I}, \ion{C}{IV}, \ion{O}{VI} and \ion{Si}{II} (top left to bottom right) for the \RT (red), the \SB \citet{mina21} (blue), and the \BW \citet{shen_2014} (green) simulation. The solid line indicates the median and the shaded region corresponds to $1\sigma$ scatter. For metal ions, the red lines indicate results where the ionisation states are calculated particle by particle using the PWPL reconstructed spectrum (from the simulation) with an extension from 500 eV to 10 keV in the {\sc Cloudy} \citep{ferland17} calculation. The red dashed line ("RT-UVB") shows results where the ionisation states are computed assuming photoionisation equilibrium with the uniform UVB from \citet{HM12} (although the element abundances, density, temperature positions etc. are from the RT simulation), similar to what has been done with the \SB and \BW simulations. \ion{H}{I} data are extracted directly from the simulation from the non-equilibrium chemical network. The observational data points from various CGM surveys are indicated with different legends. Darker symbols indicate the lowest mass halos with $\mathrm{M}_{200}\leq 10^{10.5}~\mathrm{M}_\odot$.
    }
    \label{fig:CGMprofiles}
\end{figure*}

\begin{figure}
    \centering
    \includegraphics[width=\linewidth]{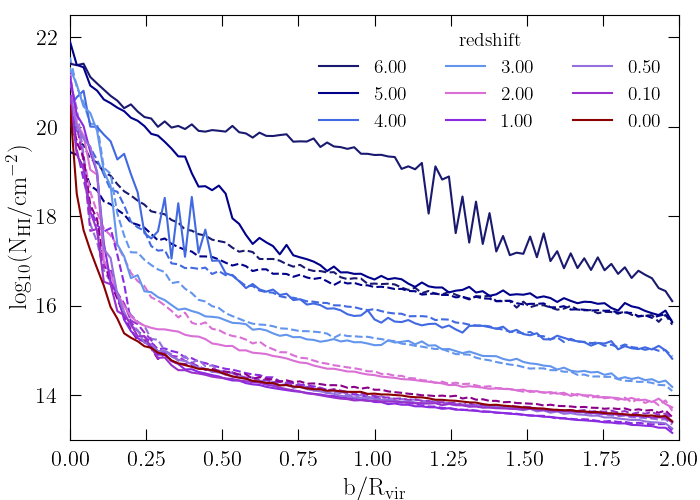}
   \caption{1D \ion{H}{I} column density distributions at different redshift ($z= 6, 5, 4, 3, 2, 1, 0.5, 0.1, 0$) as a function of impact parameter scaled with the virial radius of Bashful ($b/R_{vir}$. Solid and dashed lines correspond to the \RT and \SB run, respectively.} 
   \label{fig:HI_evo}
\end{figure}

\begin{figure}
    \centering
    \includegraphics[width=\linewidth]{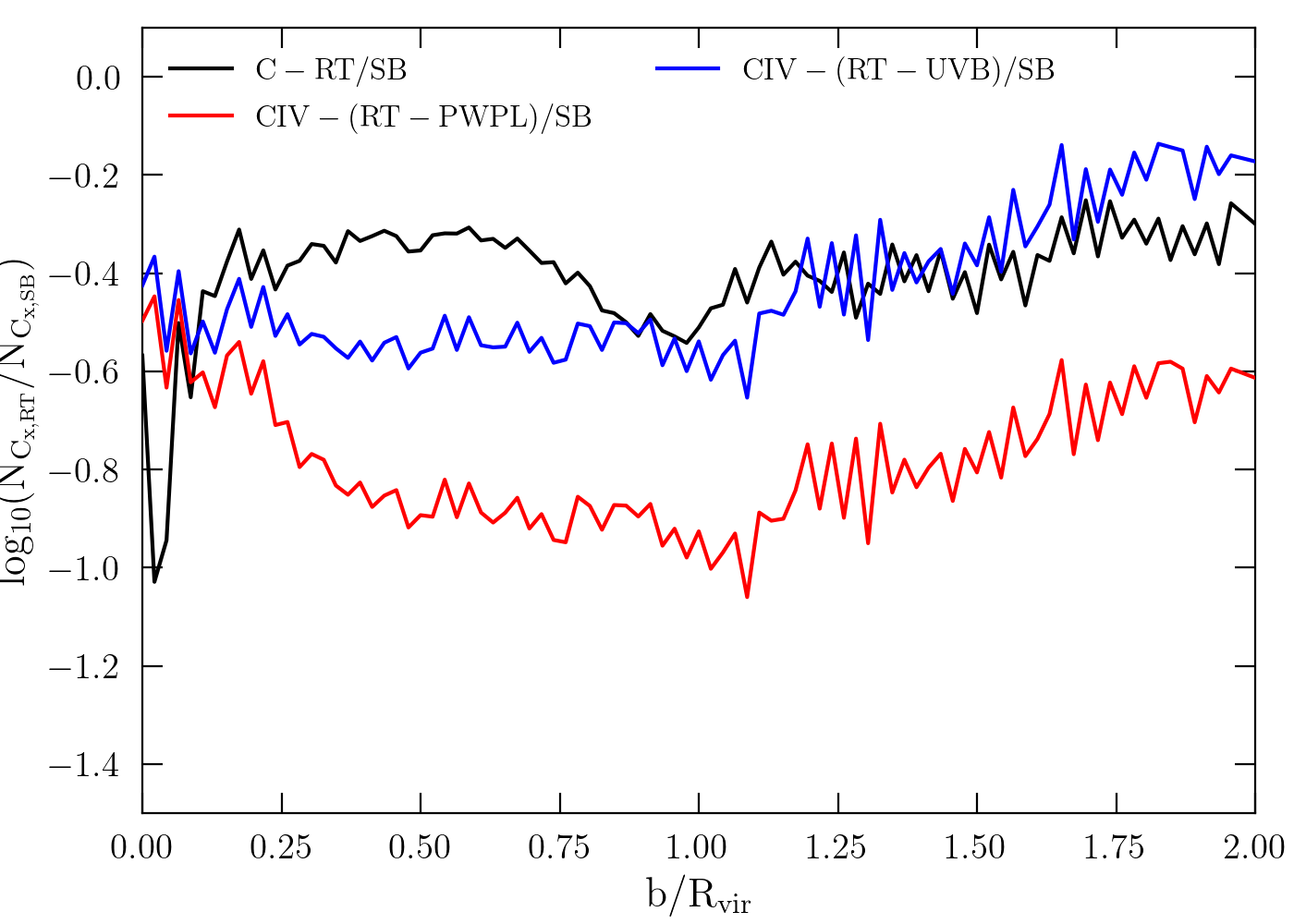}
   \caption{Column density ratios of carbon ion species between the \RT run and the \SB simulation without RT. The black line indicates the total carbon ratio. The blue line shows the ratio of \ion{C}{IV} calculated with {\sc CLOUDY} assuming a uniform extragalactic UVB. The red line indicates \ion{C}{IV} ratio of the two runs, where in the \RT run the abundance is computed using the radiation field from the simulation. }
    \label{fig:Cfraction}
\end{figure}

The circumgalactic medium contains crucial information on galaxy formation and baryonic cycles, and this is particularly important for the faintest dwarf galaxies with shallow gravitational potential wells. Although the less-bursty star formation rate (Section \ref{sec:sfr}) and smaller DM cores (Section \ref{sec:dm}) indicate somewhat weaker galactic outflows driven by SNe in our simulation compared to previous simulations without RT, the extremely low stellar metallicities at $z = 0$ (Section \ref{sec:metallicity}) still suggest that our simulated galaxies lose most of the metals through outflows into the CGM and the IGM. Indeed, at $z=0$, about 71.4\% and 91.8\% of total metals within $R_{vir}$ for Bashful and Doc are outside the central disks, respectively. 
\citet{mina21} compared the properties of the CGM between the \BW simulation and the \SB run and found that the CGM is a sensitive probe for feedback models. In this section, we investigate the impact of radiative transfer in our simulation on the CGM gas and metal distributions. 

Figure \ref{fig:MetalCGMTimeCompareSims} shows the total metal column density ($N_{z}$) at redshift $z = 6,~ 3,~ 1$ and 0 in our simulation (first row), as well as the \SB and \BW simulations (second and third row, respectively). It is clear that galactic outflows eject a substantial amount of metals into the CGM. At $z=0$, the total metal column density in the central region in our simulation appears to be smaller than that of the \SB run. However, the extent of the metal ``bubble'', defined as the impact parameter (i.e. projected distance in 2D) where the covering fraction of $N_{z} > 10^{12} \rm\, cm^{-2}$ decreases below 50\%, is 334 kpc, approximately four times the virial radius of Bashful.  
This is slightly smaller, but comparable to the metal extent in the simulation with superbubble feedback (4.7 $R_{vir}$) but much smaller than the metal extent produced in the blastwave simulation (16 $R_{vir}$). As discussed in detail in \citet{mina21}, and also in \citet{keller14}, the superbubble feedback model typically generates relatively slower winds but with a higher mass-loading factor compared to the delayed cooling blastwave approach, resulting in a less extended metal distribution. 

The fact that our RT simulation has a comparable CGM metal extent to the superbubble simulation may seem contradictory to our previous results, namely radiative feedback results in less powerful galactic outflows and the galaxies have smaller stellar masses. One may expect the CGM metals to be less spread out. However, the build-up of the metals in the CGM throughout cosmic history varies significantly between simulations. At high redshift, our RT simulation has a more extended metal distribution around each individual subhalo than the \SB simulation, and the difference is more obvious towards earlier times (see e.g. the first column of Figure \ref{fig:MetalCGMTimeCompareSims}). This is likely due to the fact that the RT simulation has predominantly early star formation, especially for the faint galaxies with $M_{h} < 10^{10} \rm\, 
 M_{\odot}$. These tiny halos have very shallow gravitational potential wells, where the galactic wind can easily pollute the surrounding CGM. It is worth noting that, by $z=3$, the CGM metal density appears to be peaked at the centres of the two main halos (the progenitors of Bashful and Doc) and monotonically decreasing outwards, but it is clear from the evolution that the contributions from the satellites are non-negligible, if not dominant, at larger distances ($\gtrsim R_{vir}$) from the galaxy centres. This is typical for high redshift systems \citep[e.g.][]{shen12}, which makes it difficult to interpret CGM observations because usually only the central, more massive galaxies are observed. 

In Figure \ref{fig:MetalCGMcompare}, we compare the column densities of commonly observed ionisation species \ion{H}{I}, \ion{C}{IV}, \ion{O}{VI} and \ion{Si}{II} between the RT simulation and the \SB simulation. In Figure \ref{fig:CGMprofiles}, we show the 1D column densities of these ions as a function of the impact parameter normalised by the virial radius ($b/R_{vir}$) of Bashful, together with recent observations of local dwarf galaxies. The \ion{H}{I} abundance is directly computed from the non-equilibrium chemistry solver coupled with radiative transfer (cf. Section \ref{sec:sim}). The abundances of metal ions are calculated particle by particle using the photonionisation code {\sc Cloudy} \citep{ferland17}, with the incident radiation spectrum from the RT simulation itself\footnote{We extended the spectrum with three additional power-law segments between 0.5-1 keV, 1-5 keV and 5-10 keV following the \citet{HM12} UVB spectrum.}. To investigate the impact of the incident radiation on ion abundances, we also calculate the abundances assuming a uniform $z=0$ UV background from \citet{HM12}, which is commonly used when post-processing simulations (including the \BW and \SB simulations) for comparison with CGM observations.  The column density along each line of sight (LOS) is calculated using the analysis tool from \citet{shen13}, by adding the contributions of particles weighted by a 2D projection of the SPH smoothing kernel.

\subsection{HI distribution}
\label{sec:HI}
For the 1D \ion{H}{I} distribution at $z=0$, on the CGM scale ($b/R_{vir} \gtrsim 0.25$), there is no significant difference between the \RT (red curves in Figure \ref{fig:CGMprofiles}) and the \SB run (blue curves). Neutral hydrogen is ubiquitous in all simulated galaxies, and rather insensitive to RT or feedback models. The median N$(\ion{H}{I})$ at $b=R_{vir}$ is about $10^{14} \rm\, cm^{-2}$, and the covering fraction ($C_{f}$) of N$(\ion{H}{I}) > 10^{13.5} \rm\, cm^{-2}$ (a common detection threshold, corresponding to an equivalent width EW $\gtrsim 100$ m\r{A}) is almost 100\% within the virial radius, which is in good agreement with observations of isolated dwarf galaxies \citep{Prochaska11, Liang14, Bordoloi14, Burchett16, Johnson17, Zheng20, Zheng24, Mishra24}. For example, \citet{Zheng24} compiled a sample of 45 isolated dwarf galaxies with the HST/COS data, and found that the \ion{H}{I} detection rate within $R_{200}$ is about 89\% in the CGM. \citet{Liang14} investigated 195 galaxy-QSO pairs and found \ion{H}{I} detection rate of 100\%. Unlike more massive galaxies, where the neutral hydrogen in the CGM exists in a clumpy structure \citep[e.g.][]{Nelson20, Decataldo24}, the \ion{H}{I} around dwarf galaxies has a rather smooth distribution. The CGM is generally highly ionised, with a \ion{H}{I} fraction ($f_{HI}$) of about $10^{-2}$ at 0.1 $R_{vir}$), which decreases to $3 \times 10^{-5}$ at the virial radius. To interpret the observational data, the authors in \citet{Zheng24} explored two empirical models for the CGM gas around a galaxy with $M_{\star} = 10^{8.3} \rm\, M_{\odot}$, both of which reproduce the observed $N(\ion{H}{I})$ distribution. Model A with lower gas density ($n_{\rm H}(r)$) and a steeper profile, but higher volume filling fraction ($f_{V} \equiv V_{cool}/V_{total}$) of cool (T$<10^{4}$ K) gas, and Model B with higher gas density and a shallower profile, but with lower volume filling fraction. The former predicts lower \ion{C}{II} (and other "low ions") column densities and higher \ion{C}{IV} columns, and the latter is the opposite. In our simulation, the volume filling fraction of the cool gas is generally not very high, ranging from 0.033 at 0.1 $R_{vir}$ to $3.4\times10^{-5}$ at $R_{vir}$, but the CGM has relatively high density $n_{\rm H}$ ranging from $7 \times 10^{-4} \rm\, cm^{-3}$ at 0.1 $R_{vir}$ to about $10^{-5} \rm\, cm^{-3}$ at $R_{vir}$. Therefore, our simulated galaxy has general CGM properties similar to Model B, although with somewhat steeper profiles for $n_{\rm H}$ and $f_{V}$.

Although the CGM \ion{H}{I} distribution is similar in all runs, at a smaller impact parameter ($b/R_{vir} \lesssim 0.25 \,R_{vir}$, i.e. the ISM and the CGM close to the disk) the \RT run has smaller \ion{H}{I} column densities. This is the case for Bashful, but particularly obvious for Doc (see Figure \ref{fig:MetalCGMcompare}), where the \ion{H}{I} in the ISM appears to be "cleared out". In fact, the total mass of \ion{H}{I} within Doc's virial radius started to decrease steadily below $z \sim 0.3$, and at $z=0$ it has dropped by an order of magnitude, with $M_{HI} = 6.7 \times 10^{5}$M$_{\odot}$ (cf. Table \ref{tab:dgs}).  Since Doc is merging into Bashful towards the end of the simulation, one may expect that ram-pressure stripping plays a role \citep{Putman21}. However, the total gas mass in Doc does not have a substantial decrease towards $z=0$, indicating that the gas is mostly photoionised by internal and external radiation. 

We emphasize that the weak dependence of CGM \ion{H}{I} distribution on supernova and radiative feedback {\it only applies to low redshift ($z\lesssim 1$).} The \ion{H}{I} distribution is drastically different between the RT run and the \SB run at high redshift. This has been illustrated in Figure \ref{fig:SigmaHI}, and we show the 1D N(\ion{H}{I}) vs. $b/R_{vir}$ at various redshifts in Figure \ref{fig:HI_evo}. Due to the fact that reionisation completes much later in the \RT simulation than in the \SB run (cf. Section \ref{sec:results}), gas remains largely neutral around the progenitor of Bashful in the \RT simulation, leading to a much higher \ion{H}{I} column of $\gtrsim 10^{20}$ cm$^{-2}$ within 1 $R_{vir}$ at $z=6$, similar to the damped Ly$\alpha$ systems (DLAs) \citep{Wolfe05}. With the completion of reionisation, the two runs start to converge at large distances, but the difference takes an opposite trend close to the galaxy at intermediate redshift ($z \sim 2-3$): the RT run has a lower N(\ion{H}{I}) than the \SB run, likely because of the radiative feedback from Bashful. The difference is largest at the inner CGM, around 0.25 $R_{vir}$, indicating a non-negligible escape fraction of ionising radiation. 

\subsection{Metal ion distributions}
The distribution of moderately ionised \ion{C}{IV} is also extended but is generally below the detection limit. Observationally, non-detection is often associated with lower-mass dwarf galaxies. For example, \citet{Burchett16} found that \ion{C}{IV} detection is more associated with galaxies with $M_{*} > 10^{9.5} \rm\, M_{\odot}$ but rarely with smaller systems. \citet{Zheng24} divided their sample of dwarf galaxies into different mass bins, and found most non-detection of \ion{C}{IV} for $M_{200} = 10^{10}-10^{10.5} \rm\, M_{\odot}$. Since Bashful has a virial mass $M_{vir} = 3.5 \times 10^{10} \rm\, M_{\odot}$ and stellar mass only $M_{*} = 4.5 \times 10^{7} \rm\, M_{\odot}$, it is not unexpected that our predictions are below the detection limit. More quantitatively, the covering fraction of N$(\ion{C}{IV}) > 10^{13.5} \rm cm^{-2}$ is only 2\% within $R_{vir}$, much below the observational findings of 22\% from \citet{Zheng24} and 23\% from \citet{Johnson17}. This may again be due to the simulated galaxy being a smaller system compared to the average halo/stellar mass in observations. The trend of decreasing \ion{C}{IV} column density in the CGM of smaller galaxies is also seen in simulations. For example, \citet{Zheng24} used the EAGLE simulation \citep{Schaye15} and went through the same analysis as the observational data and showed smaller \ion{H}{I}, \ion{C}{IV}, and \ion{C}{II} column densities in lower-mass galaxies. A similar trend is also found in \citet{Piacitelli25} with the Marvelous Dwarf and Marvelous Massive Dwarf simulations. It is clear that more statistics are needed in both observations and simulations for the CGM of the faintest dwarf galaxies $M_{\star} < 10^{8}$ M$_{\odot}$ to make more direct comparisons. 

The distribution of highly ionized \ion{O}{VI} in our simulation is more extended than \ion{C}{IV}, although it is also mostly below the detection limit for $b/R_{vir} \gtrsim 0.25$, and this is also seen in \citet{Piacitelli25} and in the FIRE-2 simulation \citep{Li21}. Observationally, the highly ionised \ion{O}{VI} is more commonly detected compared to other metal ions, and the detected \ion{O}{VI} column densities are generally higher than those in our simulation \citep{Prochaska11, Johnson17, Mishra24} even for galaxies with similar masses. Thus, our simulation underpredicts \ion{O}{VI}. As suggested in \citet{Piacitelli25}, this can be due to either not enough metals being ejected into the outer CGM, or that \ion{O}{VI}-bearing phase is underpredicted in the simulation, or both. While there is still the possibility the former occurs, the stellar metallicities provide constraints on how much metals can be ejected. Moreover, the observed kinematics of \ion{O}{VI} absorbers show that the gas is largely bound \citep{Mishra24} (at least for low redshift samples), indicating that it cannot travel to large distances. Therefore, it is more likely the oxygen is in a different ionisation state. In fact, in our simulation the CGM oxygen budget is dominated by \ion{O}{V} and \ion{O}{VII}, rather than \ion{O}{VI}. The ionisation fraction of \ion{O}{VI} increases with distance but reaches a plateau of $\sim$ 0.2 at $r > 0.6 \  R_{vir}$. This is much smaller than that of \ion{O}{VII}, which is $\sim$ 0.5-0.6 at 1 $R_{vir}$.  Moreover, the gas ionisation states depend on not only gas density and temperature, but also the radiation field. We discuss the impact of RT on CGM metal ions in Section \ref{sec:RT-CGM}.  

Unlike the \ion{H}{I} and the "high ions", the distribution of \ion{Si}{II} is very compact. The median $N(\ion{Si}{II})$ drops below $10^{13.5}$ cm$^{-2}$ only at 0.1 $R_{vir}$, indicating that \ion{Si}{II} is mainly within the ISM rather than the CGM (cf. Figure \ref{fig:MetalCGMcompare}). This is generally in good agreement with observations, which rarely detect "low ions" such as \ion{Si}{II}, \ion{Si}{III} and \ion{C}{II} in the CGM of dwarf galaxies, and this is especially true for lower mass dwarfs \citep{Johnson17, Zheng24}. As we discussed in Section \ref{sec:HI}, the CGM in our simulation is highly ionised. The gas is generally smooth and lacks the clumpy cold clouds that are often seen in the CGM of more massive galaxies \citep{Decataldo24} or in high redshift systems \citep[e.g.][]{shen13}, which give rise to low ion metal absorbers \citep[e.g.][]{Werk13, Prochaska17}.  

\subsection{Impact of RT on CGM metal distributions}
\label{sec:RT-CGM}
Compared to the \SB simulation from \citet{mina21}, our \RT simulation shows smaller column densities of \ion{C}{IV}, \ion{O}{VI}, and \ion{Si}{II}. Several factors may have contributed to this. Firstly, there are less metals in general being ejected to the CGM of Bashful and Doc, mainly because the galaxies have smaller stellar mass with radiative feedback. In particular, Bashful's stellar mass in our simulation is only about one-third of that in the \SB simulation. This effect can be seen in Figure \ref{fig:MetalCGMTimeCompareSims} ($z=0$ panels), where the total metal column $N_{Z}$ in the CGM near the galaxy is smaller in the \RT run than in the \SB run. More quantitatively, Figure \ref{fig:Cfraction} shows that the total carbon column density (black line) in the \RT run is about 0.3-0.5 dex below the \SB run, which corresponds to a 50\% to 70\% reduction in total carbon. This is consistent with the stellar mass differences in the two runs.   

In addition to suppressing star formation and consequently reducing total metal production, stellar radiation can directly affect the ionisation states of the gas and abundances of metal species. It is interesting that, if we use the radiation spectra from the radiative transfer calculation, the abundance of \ion{C}{IV} is further reduced (red line) by approximately 0.5 dex compared to the results obtained from {\sc Cloudy} calculation assuming the usual $z=0$ UVB (blue line). After investigating other species of carbon ions, we find that the most abundant ions are \ion{C}{V} and \ion{C}{VI}. The ionisation energy of \ion{C}{IV} is about 64.5 eV, which is not subject to strong absorptions from the most abundant \ion{H}{I} in the ISM \citep[e.g.][]{baumschlager23}. Thus, it appears that the hard photons that leak from the galaxy ionise \ion{C}{IV} to higher ionisation states and cause significant reduction in \ion{C}{IV} abundances, even though the star formation rate in Bashful at $z=0$ is not very high ($\sim 2 \times 10^{-3}$ M$_{\odot}$/yr). Similarly to \ion{C}{IV}, the abundances of low ion \ion{Si}{II} and high ion \ion{O}{VI} are impacted by the same effect of being ionised to higher states. For \ion{Si}{II}, the column density decreases rapidly towards a larger impact parameter, but the effect can still be seen at $b/R_{vir} \gtrsim 0.1$. The suppression of \ion{O}{VI} column density due to local radiation is generally not as prominent as for \ion{C}{IV}, most likely because the flux of X-ray photons capable of ionizing \ion{O}{VI} (138 eV) is low in stellar spectra. 

Our results show that it is important to model the ionising radiation spectrum accurately when creating synthetic observations for the CGM. In particular, photons with energy much higher than 13.6 eV can escape the ISM relatively easily and impact CGM ionisation states. The impact of local radiation on CGM observables is investigated in several recent studies, but is mostly focused on more massive galaxies. For instance, \citet{Oppenheimer18} performed zoom-in simulations from the EAGLE volume with non-equilibrium chemical network \citep{Richings14}, and found that hard radiation from flickering AGN can boost \ion{O}{VI} column densities around $L_{*}$ halos due to the long recombination timescale. More recently, \citet{Zhu24} simulated five Milky Way-like systems from the Auriga project with radiative feedback, and found that local stellar radiation dominates the radiation field within the virial radius at all redshift, and can suppress \ion{H}{I} and \ion{Mg}{II} column densities in the inner halo. The \ion{O}{VI} column density is not significantly affected by radiation due to the lack of hard photons, which is similar to what we find. However, \citet{Holguin24} performed Monte-Carlo RT on 12 FIRE simulations and concluded that at low redshift, local stellar radiation is subdominant beyond 0.2 $R_{vir}$ around MW-like halos, although it can be important at cosmic noon. As faint dwarf systems differ significantly from MW-like systems in terms of environment, assembly history, star formation history, escape fraction, and CGM properties such as density, temperature, and substructures, we cannot directly compare our results with the simulations of more massive galaxies, but we note that local radiation can potentially have stronger impact on the CGM of dwarf galaxies than to more massive systems, because the ISM of dwarf galaxies is more prone to mechanical feedback events, leading to higher escape fractions \citep[e.g.][]{Kimm14}. 

Of course, the impact of radiation on the CGM is not only on the observables; the changing of the ionisation states of metal ions would fundamentally alter the metal cooling rate, and consequently the dynamics of gas accretion and galaxy evolution \citep[e.g.][]{Obreja24}. Ideally, simulations should incorporate on-the-fly RT coupled with non-equilibrium chemical solvers for metal ions in addition to the primordial species \citep[e.g.][]{Katz22}, but this approach can be computationally prohibitive for cosmological simulations that evolve to $z=0$. Alternatively, one can assume ionisation equilibrium, and use pre-tabulated tables with varying radiation field parameters \citep{Gnedin12}. We defer a more detailed study on the impact of local radiation field on metal cooling and gas dynamics in a future work.

\section{Conclusions}\label{sec:summary}

We present results from a cosmological zoom-in simulation of a small group of field dwarf galaxies. The simulation is performed with the {\sc Gasoline2} code coupled with the on-the-fly radiative transfer algorithm {\sc Trevr2} \citep{wadsley23} and evolved to $z=0$. Stellar radiation is included according to an age-dependent spectral energy distribution calculated from {\sc Starburst99} \citep{leitherer14}, and a redshift-dependent UV background radiation \citep{HM12} is modelled with background particles, and is turned on at $z=15.1$ and propagates toward the simulation volume. The ionisation, heating and cooling of primordial gas are computed with non-equilibrium chemical network and coupled with RT, but radiative cooling from metal lines are calculated with {\sc Cloudy} table assuming photoionisation equilibrium with the uniform UVB. The initial condition, resolution, and other galaxy formation physics such as star formation, metal enrichment, and supernova feedback are identical to the Seven Dwarf simulation by \citet{mina21}, which we call \SB run in comparison. We also compare with the first version of the Seven Dwarf simulation from \citet{shen_2014}, which uses delayed cooling model for SN feedback, and we refer to it as the \BW run. The main focus of this work is to investigate the impact of radiation in dwarf galaxy formation and the interplay between radiative feedback, star formation, and supernova feedback in shaping galaxy properties and the surrounding CGM.   

There are in total 12 galaxies formed within our simulation volume: 1) two luminous dwarf galaxies, Bashful and Doc, which reside in halos with $M_{vir} > 10^{10} \rm\, M_{\odot}$. They are gas-rich and star-forming towards $z=0$, although Doc's ISM is largely photoionised by internal and external radiation towards low redshift. 2) Two faint dwarf galaxies, Dopey and Grumpy, formed within halos of mass $10^{9} \rm\, M_{\odot} < M_{vir} < 10^{10} \rm\, M_{\odot}$. They have stellar masses of several $10^{5} \rm\, M_{\odot}$ and completed star formation by $z=3.2$, and thus have only old stellar populations. These 4 galaxies also formed in the original \BW simulation. 3) An additional population of 8 quenched galaxies that have all accreted into Bashful's halo by $z=0$, which we label as qDG1-8. They reside in halos with masses varying from $10^{6} \rm\, M_{\odot}$ to several $10^{8} \rm \,M_{\odot}$ and have typical stellar masses of $10^{4} \rm\, M_{\odot}$ to a few times $10^{5} \rm\, M_{\odot}$. Their DM halos are strongly stripped during the accretion to Bashful. Most of the star formation occurred at $z \sim 4$. These qDGs, together with Dopey and Grumpy, have luminosities and star formation histories similar to the observed Ultra-Faint Dwarf galaxies (UFDs) in the Local Group. 

The main findings of this work are summarised as follows: 
\begin{itemize}
\item The formation of Dopey, Grumpy and qDGs is strongly influenced by cosmic reionisation. The main reason these galaxies have early star formation is that the ionizing background radiation requires a finite timescale to propagate into the simulation volume. Although UVB is "turned on" at $z=15.1$, the IGM in the simulated volume is not fully ionised before $z \sim 4$, leaving enough time for gas within the smallest halos to remain cold, condense, and form stars. This propagation time delay is captured in our \RT run, but not in the previous \SB and \BW simulations, where the uniform UVB is turned on instantaneously within the whole simulation region. 
\item Radiative feedback decreases the total stellar mass of the two most massive galaxies, Bashful and Doc, by a factor of 2-3. It also results in less bursty star formation rates (SFR) than in the \SB simulation. Similar to other studies with RT, photoionisation heating preconditions the ISM such that star formation is less clustered, which in turn results in weaker collective SN feedback and outflows. The SFH of both classic and ultra-faint dwarf galaxies are consistent with observations.  

\item The simulated galaxies have many properties which are generally consistent with observed scaling relations, including the SMHM ($M_{*}-M_{h}$) relation, mass-size ($M_{*}-r_{1/2}$) relation, and luminosity-velocity dispersion ($M_{V}-\sigma_{los}$) relation. Compared to the \SB simulation, for Bashful and Doc, radiative feedback reduces the stellar disc sizes, possibly because of the earlier star formation histories and reduced galactic outflows. The velocity dispersion is not strongly impacted by RT, but dynamic interactions and tidal stripping significantly reduce the stellar velocity dispersion of qDGs. 

\item The simulated dwarf galaxies are extremely metal poor, with average [Fe/H] ranging from -3.3 to -1.5. Although the stellar metallicity (indicated by [Fe/H]) of Bashful and Doc are in good agreement with the observed mass/luminosity-metallicity relation (MZR), the stellar metallicity of the faint dwarf galaxies (Dopey, Grumpy and the qDGs) with V-band luminosity $L_{V} < 10^{6} \rm\, L_{\odot}$ is systematically lower than the observations of UFDs. The metallicity distribution function (MDF) of these galaxies peaks at [Fe/H] $\sim -3.5$, which is about an order of magnitude lower than observations from \citet{Fu23}. This may indicate that our SN feedback ejects too much metals but can also be due to the uncertainties in iron yields from both Type Ia and Type II supernova at high redshift. The actual average metallicity ($Z/Z_{\odot}$) of the faint dwarf galaxies are more consistent with the MZR.  

\item At $z=0$, Bashful and Doc have dark matter cores of size about 1.12 kpc and 0.96 kpc, respectively, which are about 3(2) times smaller than those in the \SB run for Bashful (Doc). The average central DM densities within 150 pc are $5.8 \times 10^{7} \rm\, M_{\odot} \rm kpc^{-3}$ and $4.7 \times 10^{7} \rm\, M_{\odot} \rm kpc^{-3}$, respectively, which are in agreement with data from \citet{read19b}. Compared to the \SB simulation, the baryonic mass fluctuation at the galaxy centres ($<$500 pc) are milder due to less impulsive galactic winds and preventive feedback from local stellar radiation. This leads to less fluctuation in the central gravitational potential well and consequently less DM heating, and smaller cores. DM core formation starts early, but unlike previous simulations, the core size growth is rather smooth and without abrupt changes that are associated with rapid outflows and inflows. 
\item Among the fainter dwarf galaxies, only Dopey, Grumpy and qDG2 have marginally resolved DM cores with $r_{c}=$ 0.29 kpc, 0.34 kpc and 0.49 kpc, respectively. These three galaxies have stellar masses of several $10^{5} \rm \,M_{\odot}$, which are among the most massive ones in the faint galaxy sample. Our results are in agreement with some higher resolution UFD simulation where SN feedback is resolved \citep[e.g.][]{agertz20} but not all. Moreover, it is interesting that qDG3 does not have a resolved DM core, despite that it is more massive than the 3 galaxies which formed cores. Since this galaxy is quenched early, it suggests that dynamic interactions such as mergers can have non-negligible impact on DM cusp-core transformation. 

\item The extent of the metal-enriched volume in the CGM in our \RT simulation is comparable to that in the \SB simulation at $z=0$, about four times of the virial radius of Bashful. This confirms previous findings that dwarf galaxies are very efficient CGM polluters. However, due to the early star formation history in our simulation (before reionisation is completed), at high redshift ($z\gtrsim 3$) the CGM metal extent is much larger than the ones in previous Seven Dwarf simulations. Even at $z=0$, the contribution from satellite progenitors to the CGM metals at large distances are non-negligible. 

\item The \ion{H}{I} distribution on the CGM scale (with impact parameter $b \gtrsim 0.25 \,R_{vir}$) is ubiquitous and rather insensitive to RT or feedback models. At $R_{vir}$, the median column density N(\ion{H}{I}) is about $10^{14} \rm\, cm^{-2}$ with unity covering fraction, which is in good agreement with observations. The CGM is highly ionised, with relatively low volume filling fraction of cold, neutral gas. At the inner halo or ISM scale, however, local stellar radiation decreases \ion{H}{I} column densities. In particular, the impact of photoionsation heating combined with ram-pressure stripping causes Doc to lose majority of its ISM close to $z=0$. 

\item The distributions of high ions such as \ion{C}{IV} and \ion{O}{VI} are extended, whereas low ions such as \ion{Si}{II} are very compact which declines sharply beyond the ISM scale ($> 0.1 \,R_{vir}$). Nevertheless, the metal column densities are generally below the observational detection limit, especially at larger distances. The covering fractions within the virial radius are also smaller compared to observations. However, it is unclear whether there is a true discrepancy, or due to the fact that the observation data are for more massive dwarf galaxies, with the median stellar mass being an order of magnitude larger than Bashful. 

\item Compared to the \SB simulation, our simulation resulted in smaller column densities of \ion{C}{IV}, \ion{O}{VI} and \ion{Si}{II}. The main reasons are: 1) The total metal production is reduced because radiative feedback suppresses star formation in Bashful and Doc; 2) Photons with energies higher than hydrogen ionisation can escape the ISM and further ionise metal ions in the CGM, in particular the moderately ionised \ion{C}{IV}, which is reduced by approximately 0.5 dex compared to the results obtained from {\sc Cloudy} calculations assuming a uniform UVB. Thus, it is important to model the ionising radiation spectrum accurately when creating synthetic CGM observations.

\end{itemize}

These findings demonstrate that radiative transfer significantly influences dwarf galaxy formation, regulating star formation, feedback processes, and the emergence of observable properties.
Our results highlight the need for future simulations to incorporate full radiative feedback to accurately model the formation of faint galaxies in the early universe.

\begin{acknowledgements}
  BB and SS acknowledge the support of the Research Council of Norway through the NFR Young Research Talents Grant 276043.     
  SS and RW acknowledge support from the European High Performance Computing Joint Undertaking (EuroHPC JU) and the Research Council of Norway through the funding of the SPACE Center of Excellence (grant agreement N0 101093441). Parts of the computations were performed on resources provided by Sigma2 - the National Infrastructure for High-Performance Computing and Data Storage in Norway. BWK acknowledges support from the Space Telescope Science Institute, which is operated by the Association of Universities for Research in Astronomy, Inc. under NASA contract NAS 5-26555. This funding is associated with HST Theory Program AR-17547. A.M.B acknowledges support from NASA Grant 80NSSC24K0894 and FI-CCA-Research-00011826 from the Simons Foundation.
\end{acknowledgements}

\bibliographystyle{aa}
\bibliography{7dwarfs}
\end{document}